\documentclass{article}

\usepackage{microtype}
\usepackage{graphicx}
\usepackage{subfigure}
\usepackage{booktabs} 

\usepackage{hyperref}

\usepackage[accepted]{icml2024}

\usepackage{amsmath}
\usepackage{amssymb}
\usepackage{mathtools}
\usepackage{amsthm}

\usepackage[capitalize,noabbrev]{cleveref}

\theoremstyle{definition}

\theoremstyle{remark}

\usepackage{algorithm}
\usepackage{algorithmic}
\usepackage{tikz}
\usepackage{amssymb}
\usepackage{amsthm}
\usepackage{amsmath}
\usepackage{mathabx}
\usepackage{mathtools}
\usepackage{dsfont}
\usepackage{listings}
\usepackage{float}
\usepackage{tcolorbox}
\usetikzlibrary{patterns,automata,positioning,arrows}
\usepackage{bm}
\usepackage{enumitem}
\usepackage{microtype}

\usepackage{array}
\newcolumntype{x}[1]{>{\centering\arraybackslash\hspace{0pt}}p{#1}}

\global\long\def\e{\mathbf e}

\global\long\def\eta{1_{\left|\mathcal{A}_{1}\right| \times 1}^\top}

\global\long\def\I{\mathcal I}
\global\long\def\J{\mathcal J}
\global\long\def\mA{\mathcal{A}}
\global\long\def\mAa{\mathcal{A}_{1}}
\global\long\def\mAb{\mathcal{A}_{2}}

\global\long\def\mR{R}

\global\long\def\p{\mathbf p}
\global\long\def\pia{\p}

\global\long\def\pib{\q}

\global\long\def\q{\mathbf q}
\global\long\def\r{\mathbf r}
\global\long\def\R{\mathbb R}
\global\long\def\s{\mathbf s}
\global\long\def\sc{\textup{SC}}

\global\long\def\st{\textup{s.t.}}

\global\long\def\mQ{\mathds{Q}}
\global\long\def\mS{\mathcal{S}}

\newtheorem{thm}{Theorem}

\newtheorem{lem}[thm]{Lemma}
\newtheorem{prop}[thm]{Proposition}

\newtheorem{df}{Definition}
\newtheorem{eg}{Example}

\newtheorem{cond}{Condition}
\newtheorem{rmk}{Remark}

\usepackage[textsize=tiny]{todonotes}

\icmltitlerunning{Minimally Modifying a Markov Game to Achieve Any Nash Equilibrium and Value}

\begin{document}

\twocolumn[
\icmltitle{Minimally Modifying a Markov Game to Achieve Any Nash Equilibrium and Value}



\icmlsetsymbol{equal}{*}

\begin{icmlauthorlist}
\icmlauthor{Young Wu}{uwmcs}
\icmlauthor{Jeremy McMahan}{uwmcs}
\icmlauthor{Yiding Chen}{corn}
\icmlauthor{Yudong Chen}{uwmcs}
\icmlauthor{Xiaojin Zhu}{uwmcs}
\icmlauthor{Qiaomin Xie}{uwmis}
\end{icmlauthorlist}

\icmlaffiliation{uwmcs}{Department of Computer Sciences, University of Wisconsin--Madison, Madison, Wisconsin, United States}
\icmlaffiliation{uwmis}{Department of Industrial and Systems Engineering, University of Wisconsin--Madison, Madison, Wisconsin, United States}
\icmlaffiliation{corn}{Department of Computer Science, Cornell University, Ithaca, New York, United States}

\icmlcorrespondingauthor{Young Wu}{yw@cs.wisc.edu}

\icmlkeywords{Multi-agent Reinforcement Learning, Game Theory, Adversarial Machine Learning, ICML}

\vskip 0.3in
]



\printAffiliationsAndNotice{}  
\allowdisplaybreaks
\begin{abstract}
We study the game modification problem, where a benevolent game designer or a malevolent adversary modifies the reward function of a zero-sum Markov game so that a target deterministic or stochastic policy profile becomes the unique Markov perfect Nash equilibrium and has a value within a target range, in a way that minimizes the modification cost. We characterize the set of policy profiles that can be installed as the unique equilibrium of a game and establish sufficient and necessary conditions for successful installation. We propose an efficient algorithm that solves a convex optimization problem with linear constraints and then performs random perturbation to obtain a modification plan with a near-optimal cost. The code for our algorithm is available at: \url{https://github.com/YoungWu559/game-modification}.
\end{abstract}

\section{Introduction}

Consider a two-player zero-sum Markov game $G^\circ=(R^\circ, P^\circ)$ with payoff matrices $R^\circ$ and transition probability matrices $P^\circ$. Let $\mathcal{S}$ be the finite state space, $\mA_i$  the finite set of actions for player $i\in\{1,2\}$, and $H$ is the horizon.
It is well known that such a game has at least one Markov Perfect (Nash) Equilibrium (MPE) $(\p^\circ,\q^\circ)$, where $\p^\circ$ is the Markov policy for player 1 and $\q^\circ$ for player 2 \citep{maskin2001markov}.
Furthermore, all the MPEs of $G^\circ$ have the same game value $v^\circ$, corresponding to the expected payoff for player 1 and loss for player 2 at equilibrium. In the special case with $H=1$ stage, the Markov game reduces to a matrix normal form game and the MPE reduces to a Nash Equilibrium (NE).

There may be reasons for a third party to prefer an outcome with a different MPE and/or game value.
See Appendix~\ref{app:water} for a concrete example. 
A \textbf{benevolent} third party may want to achieve fairness.
Many games are unfair in that $v^\circ \neq 0$ (an example, two-finger Morra, is given in the experiment section).  The third party can modify the payoffs $R^\circ$ into $R$ such that the new game given to the players is fair with value $v=0$.
Similarly, many games have non-intuitive MPEs, and players with bounded rationality (e.g., average people) may fail to find or implement them. For the benefit of such players, the third party may seek a new game whose MPE $(\p,\q)$ is an intuitive strategy profile, such as uniform randomization among a set of actions.

In addition, one often desires an MPE consisting of stochastic policies (i.e., a \emph{mixed} strategy equilibrium). If actions represent resources (roads, advertisement slots, etc.), the game designer might want all resources to be utilized; if actions represent customers, requests, or demands, the designer might want all of them to be served; if a board/video game is concerned, the designer might want the agents to take diverse actions so that the game is more entertaining. 
Conversely, a \textbf{malicious} third party may want to trick the players into playing an MPE $(\p, \q)$ of its choice. As most games have mixed equilibria, the players may get suspicious if the modified game turns out to have a pure strategy MPE, whereas a mixed equilibrium is harder to detect.
Furthermore, the adversary may want to control the game value $v$ to favor one player over the other---this is the analog of adversarial attacks in supervised learning.

Regardless of intention, game modification typically incurs a cost to the third party, who seeks to minimize it. We assume that the cost is measured by some loss function $\ell(R,R^\circ)$ depending on the new and original games $R$ and $R^\circ$. For example, one may consider $\ell(R,R^\circ)=\|R-R^\circ\|$ for some norm $\|\cdot\|$. 

\paragraph{The Game Modification Problem}

It is important to understand when efficient modification is possible and to understand malicious attacks so as to develop an effective defense. This motivates us to study the following \emph{Game Modification} problem, specified by the tuple $$(R^\circ, P^\circ, b, (\p,\q), [\underline{v}, \overline{v}], \ell).$$
Here $R^\circ$ and $P^\circ$ are the payoff and transition matrices, respectively, of the original Markov game.
A valid payoff value must be in $[-b,b]$. 
The third party has in mind an arbitrary (and potentially stochastic) target MPE $(\p,\q)$,
which is typically not the unique MPE of $R^\circ$.
The third party also has in mind a target game value range $[\underline{v}, \overline{v}]$. It is possible that $b=\infty$, $\underline{v}=-\infty$ or $\overline{v}=\infty$.

\begin{df} [Game Modification] \label{def:GM}

Game modification is the following optimization problem to find $R$ given $(R^\circ, P^\circ, b, (\p,\q), [\underline{v}, \overline{v}], \ell)$:
\begin{eqnarray}
\label{eq:GM}
\inf_R && \ell(R,R^\circ) \\
\mbox{s.t.} && (\p,\q) \mbox{ is the unique MPE of } (R, P^\circ) \nonumber \\
	&& \mathrm{value}(R, P^\circ) \in [\underline{v}, \overline{v}], \;\;
    R \mbox{ has entries in } [-b, b]. \nonumber
\end{eqnarray}
\end{df}
It is important to require that the modified game $(R, P^\circ)$ has a \textbf{unique} MPE.
In this case, no matter what solver the players use, they will inevitably find $(\p,\q)$ and not some other MPEs of $R$.
Henceforth, we refer to a Markov game simply by its payoff matrices $R$ and suppress reference to the transition matrices $P^\circ$, which the third party cannot change.

To the best of our knowledge, the Game Modification problem in the generality of Definition~\ref{def:GM} has not been studied in the literature.
With a potentially mixed target MPE $(\p,\q)$ and the constraints on uniqueness, game value, and payoffs, it is a priori unclear when the optimization problem~\eqref{eq:GM} has a feasible solution. Moreover, in addition to just finding one feasible game or checking the feasibility of a specific game, we need to solve the harder problem of \emph{optimizing} over all feasible games with a target strategy as the unique NE.  The multi-step structure of Markov games further complicates the problem.

\paragraph{Our Contributions}

In this paper, we answer the above questions: we provide a sufficient and necessary condition for the feasibility of the game modification problem and develop an efficient algorithm that provably finds a near-optimal solution under convex losses $\ell$. In particular, using an operational characterization of MPE uniqueness, we formulate the game modification problem as an optimization problem with linear and spectral constraints and completely characterize its feasibility. We further propose an efficient Relax and Perturb algorithm circumventing the spectral constraint's nonconvexity and establish the algorithm's correctness and near-optimality.

We first study the special case of normal form games in Section~\ref{sec:normalformgame}, followed by a generalization to Markov games in Section~\ref{sec:markovgame}.

\section{Related Work} 
\label{sec:related}

Reward modification in single-agent reinforcement learning has been studied in \citet{banihashem2022admissible, huang2019deceptive, rakhsha2021policy, rakhsha2021reward, rakhsha2020policy, zhang2020adaptive}. In this setting, a deterministic optimal policy always exists. Generalizing to the multi-agent setting, even in the zero-sum case, involves the complication of multiple equilibria and the non-existence of deterministic equilibrium policies.

Adversarial attacks on multi-agent reinforcement learners are studied in~\citet{wu2023reward, ma2021game}, who consider the setting where an attacker installs a target \emph{dominant strategy equilibrium} by modifying the underlying bandit or Markov game. In general, mixed strategies that assign positive probabilities to multiple actions cannot be dominant (they are not dominated by at least one of the actions in the support). Therefore, the approach in~\citet{wu2023reward, ma2021game} cannot be directly applied in our setting, which targets at a mixed strategy Nash equilibrium. 

Our model is similar to~\citet{wu2023faking}, where an attacker installs a target Nash equilibrium by poisoning the training data set. Their work requires the target equilibrium to be a \emph{deterministic} action profile (i.e., not mixed), and they assume the victims estimate confidence regions of the game payoff matrices based on a noisy data set. Since it is generally impossible for all games in the confidence region to have the same mixed strategy Nash equilibrium, the modification goal in our setting is infeasible under their setting. Similarly, data poisoning techniques in~\citet{ma2019policy, rangiunderstanding, zhang2008value, zhang2009policy} do not apply to our setting. 
Instead, we consider the problem in which the players are provided with the exact payoff matrix by the game designer, so it is possible to install a mixed strategy as the unique equilibrium of the modified game. 
\citet{monderer2003k, anderson2010internal} explore the problem of installing a pure strategy equilibrium while minimizing the modification cost, but their method does not directly extend to mixed-strategy equilibria. 

As our work concerns \emph{optimizing} over the set of games with a target strategy as the unique NE, we need a \emph{sufficient and necessary} condition for uniqueness. Related and partial results can be found in a line of prior work on matrix games with a unique Nash equilibrium~\citep{kreps1974bimatrix, millham1972constructing, heuer1979uniqueness, quintas1988uniqueness,discrete1950hf} and the related problem of unique optimal solutions to linear programs~\citep{mangasarian1978uniqueness, appa2002uniqueness, szilagyi2006uniqueness}.  Our work gives a form of sufficient and necessary condition that is amenable to being used as constraints in a cost minimization formulation, and we provide a short proof. We also go beyond prior work by studying when this condition is satisfiable under the additional value and payoff constraints in~\eqref{eq:GM}, with generalization to Markov games.


\section{Modifying Normal Form Games}
\label{sec:normalformgame}

We begin with the game modification problem for matrix normal form games, which is a special case of Markov Games with horizon $H=1.$

\subsection{Preliminaries}

Consider a finite two-player zero-sum game with action space $\mA = \mAa \times \mAb$ and a $b$-bounded payoff matrix $R \in [-b,b]^{|\mAa| \times |\mAb|}$.
When a joint action $(i,j)\in \mAa \times \mAb$ is played, player 1 receives reward
 $[\mR]_{i j}$ and player 2 receives reward $-[\mR]_{i j}$. 
Let $(\p, \q)$ denote a (possibly mixed) strategy profile, where $\p \in \Delta_{\mAa}$ and $\q \in \Delta_{\mAb}$, with $\Delta_{\mathcal{D}}$ denoting the probability simplex on $\mathcal{D}$. The expected reward for player 1 is given by $\p^\top R \q.$

NE can be defined in several equivalent ways. Most convenient for us is the following definition in terms of lack of incentive for unilateral deviation. A finite two-player game has at least one NE and possibly more~\citep{nash1950equilibrium}.
\begin{df} [Nash Equilibrium] \label{df:ne} 
$(\p,\q)$ is a Nash Equilibrium of a game $R$ if 
and only if $\p^\top R \q \geq \p'^\top R \q$ for all $ \p' \in \Delta_{\mAa}$ and $\p^\top R \q \leq \p^\top R \q'$ for all $ \q' \in \Delta_{\mAb}.$
\end{df}

\subsection{Equivalent Formulation of Game Modification}

As stated in Definition~\ref{def:GM}, the game designer seeks a least-cost game with a given $(\p,\q)$ as the unique NE and satisfying the value and payoff constraints in \eqref{eq:GM}. 
To understand when such a game exists and how to find the optimal game algorithmically, our first step is to provide an equivalent formulation where the uniqueness requirement is expressed explicitly as linear and spectral constraints. 

This is done in the theorem below, for which some notations are needed. Let $\I=supp(\p)$ and $\J=supp(\q)$ denote the supports. We use  $\left[R\right]_{\I \J}$ or $R_{\I \J}$ to denote the $\left| \I \right| \times \left| \J \right|$ submatrix of $R$ with rows in $\I$ and columns in $\J$. We write $R_{\I \bullet}$ for the $\left| \I \right| \times \left| \mAb \right|$ submatrix with rows in $\I$, and $R_{\bullet \J}$ for the $\left| \mAa \right| \times \left| \J \right|$ submatrix with columns in $\J$.
Denotes by ${\bm 1}_{|\I|}$ the $|\I|$-dimensional all-one vector. 

\begin{prop} 
[Reformulation of Normal-Form Game Modification] 
\label{thm:gmnf} 
For normal form games and a target policy $\left(\p, \q\right)$ with supports $\I, \J$, the game modification problem~\eqref{eq:GM} is equivalent to the following optimization problem:
\begin{subequations} 
\label{eq:GMO}
\begin{align}
\displaystyle\inf_{R,v} \; & \;\ell\left(R, R^\circ\right)
\\ \st\; & R_{\I\bullet} \q = v {\bm 1}_{\left| \I \right|} && \textup{[row SII]} \label{eq:inqsame}
\\ & \p^\top R_{\bullet\J} = v {\bm 1}^\top_{\left| \J \right|} && \textup{[column SII]} \label{eq:inpsame}
\\ & R_{\mAa \setminus \I \bullet} \q < v {\bm 1}_{\left| \mAa \setminus \I \right|} && \textup{[row SOW]} \label{eq:outqworse}
\\ & \p^\top R_{\bullet \mAb \setminus \J} > v {\bm 1}^\top_{\left| \mAb \setminus \J \right|} && \textup{[column SOW]} \label{eq:outpworse}
\\ & \sigma_{\min}\!\!\left(\begin{bmatrix} R_{\I \J} \!\!&\!\! -{\bm 1}_{\left| \I \right|} \\ {\bm 1}^\top_{\left| \J \right|} & 0 \end{bmatrix}\!\right) \!\!>\! 0 && \textup{[INV]} \label{eq:INV}
\\ & \underline{v} \leq v \leq \overline{v} && \textup{[value range]} \label{eq:GMO_value_range}
\\ & -b \!\leq\! R_{ij} \!\leq\! b, \forall(i,j)\in \mA && \textup{[payoff bound]} \label{eq:GMO_payoff_bounds}
\end{align} 
\end{subequations}
where $\sigma_{\min}(\cdot)$ denotes the smallest singular value.
\end{prop}

Proposition~\ref{thm:gmnf} follows immediately from the lemma below, which shows that SIISOW and INV constitute a sufficient and necessary condition for a game $R$ to admit a given $(\p,\q)$ as the unique NE.

\begin{lem} [Uniqueness of NE] \label{lem:unique} 
$R$ has a unique Nash equilibrium $\left(\p,\q\right)$ if and only if $R$ satisfies both SIISOW (Condition~\ref{def:gap}) and INV (Condition~\ref{df:inv}) with respect to $(\p,\q)$:
\end{lem}

\begin{cond}[SIISOW: Switch-In Indifferent, Switch-Out Worse]
\label{def:gap}
A game $R$ satisfies SIISOW with respect to $(\p, \q)$ if equations \eqref{eq:inqsame}, \eqref{eq:inpsame}, \eqref{eq:outqworse} and \eqref{eq:outpworse} hold.
\end{cond}

\begin{cond}[INV: Invertability]
\label{df:inv}
A game $R$ satisfies INV with respect to $(\p, \q)$ if  
equation~\eqref{eq:INV} holds, 
that is, the matrix $\begin{bmatrix} R_{\I\J} & -{\bm 1}_{|\I|} \\ {\bm 1}_{|\J|}^\top & 0 \end{bmatrix}$ is invertible.
\end{cond}

If the strict inequalities in SIISOW were changed to weak inequalities, the four equations would be equivalent to Definition~\ref{df:ne} of NE \citep{osborne2004introduction}. Therefore, SIISOW implies that $(\p,\q)$ is an NE of $R$. Moreover, under this NE, if the other player switches to any pure strategy outside its NE support, its reward will be \emph{strictly} worse by equations~\eqref{eq:outpworse} and \eqref{eq:outqworse} (``switch-out worse''); 
if the other player uses any pure strategy within its support, it will achieve the same game value by equations~\eqref{eq:inpsame} and~\eqref{eq:inqsame} (known as the ``switch-in indifference'' principle). 


We are not aware of an NE uniqueness result stated in this form in the literature, though several partial and related results exist. Lemma C.3 in \citet{mertikopoulos2018cycles} implies that the SIISOW condition is necessary for NE uniqueness. Several papers study the existence of (or explicitly construct) a game with a unique NE~\cite{kreps1974bimatrix,millham1972constructing,discrete1950hf,nagarajan2020chaos}; our Lemma~\ref{lem:unique} characterizes \emph{all} such games and thereby allow one to optimize over them, as done in the formulation~\eqref{eq:GMO}. In Appendix~\ref{app: unique}, we provide a short, self-contained proof for Lemma~\ref{lem:unique}, noting that with some additional work one may also derive the lemma from the results in \citet{szilagyi2006uniqueness} on unique solutions to linear program (LP). We remark that \citet{appa2002uniqueness, mangasarian1978uniqueness} also provide uniqueness results for LP, but they are in terms of perturbation stability of the solution
and hence not in an operational form that can be used as constraints in a cost-minimization problem like~\eqref{eq:GMO}.


\subsection{Feasibility of Game Modification}

We now study when Game Modification in normal-form games, as formulated in Proposition~\ref{thm:gmnf}, is feasible. The following theorem provides a \emph{sufficient and necessary} condition. 


\begin{thm} [Feasibility of Game Modification] \label{thm:feasible} 
The Game Modification problem in Proposition~\ref{thm:gmnf} 
for normal-form games is feasible if and only if $(\p,\q)$ satisfies $\left| \I \right| = \left| \J \right|$ and it holds that $\left(-b, b\right) \cap \left[\underline{v}, \overline{v}\right] \neq \emptyset$.

\end{thm}
The equal-support condition $\left| \I \right| = \left| \J \right|$ arises due to the INV condition, which requires $R_{\I\J}$ to be a square matrix. The necessity of the equal-support condition is known in \citet{kreps1974bimatrix, millham1972constructing, heuer1979uniqueness}. Our lemma further establishes the necessity of the condition $\left(-b, b\right) \cap \left[\underline{v}, \overline{v}\right] \neq \emptyset$. Note that the game value cannot equal $b$ or $-b$, because the SIISOW condition stipulates a strictly positive gap between the game value and the value of the off-support actions. The complete proof is provided in the Appendix~\ref{app: feas}. 

The other direction of our proof is constructive.
We present a special matrix game called Extended Rock-Paper-Scissors (eRPS), which has the desired $(\p,\q)$ as the unique NE. This game can be defined for arbitrary strategy space sizes $|\mAa|$ and $ |\mAb|$. The standard rock paper scissors game is a special case when the sizes are 3, hence the name. 

\begin{df} [Extended Rock-Paper-Scissors Game] 
\label{df:erps} 
Given strategy spaces $\mAa, \mAb$, and target strategy profile
$\left(\p, \q\right) \in \Delta_{\mAa}\times \Delta_{\mAb}$
with equal supports $\I=\J=\{0,\ldots,k-1\}$, where $1 \le k \le \min(|\mAa|,|\mAb|)$,
the Extended Rock Paper Scissors Game $R^{\textup{\;eRPS\;}(\p,\q)}$ is:
\begin{equation} \begin{aligned}
& R^{\textup{\;eRPS\;}(\p,\q)}_{ij} = 
\begin{cases} 
- \dfrac{c}{\p_i \q_j} & \text{\;if\;} \substack{ k > 1, i,j < k \\ j = (i + 1) \mod k \\ } \\ 
\dfrac{c}{\p_i \q_j} & \text{\;if\;} \substack{ k > 1, i,j < k \\ j = (i + 2) \mod k \\ } \\ 
1 & \text{\;if\;} i < k, j \geq k \\ 
-1 & \text{\;if\;} i \geq k, j < k \\ 
0 & \text{\;otherwise\;}, \\ \end{cases}
\end{aligned} \end{equation}
where 
$
c = \min_{i \in \I} \left(\p_{i} \q_{(i + 1 \mod k)}, \p_{i} \q_{(i + 2 \mod k)}\right)
$
is a normalizing constant ensuring that all the entries of $R^{\textup{\;eRPS\;}}$ are between $-1$ and $1$.
\end{df}

For support size $k=1$, namely $(\p,\q)$ is a pure strategy profile, the $R^{\textup{\;eRPS\;}}$ game is visualized in Table~\ref{tab:eRPS1} (left).  It is easy to check that the upper left corner $(0,0)$ is indeed the unique pure Nash equilibrium.

For support size $k\ge 2$, namely $(\p,\q)$ is a mixed strategy profile, the $R^{\textup{\;eRPS\;}}$ game is visualized in Table~\ref{tab:eRPS1} (right) and Table~\ref{tab:eRPS}. As a special case, for $\p=\q=(1/3, 1/3, 1/3)$, $R^{\textup{\;eRPS\;}}$ is the standard Rock-Paper-Scissors game.


\begin{table}[ht]
\begin{center} 
\begin{minipage}[c]{.35\linewidth}
{\small
\setlength\tabcolsep{3pt} 
\begin{tabular}{|x{.45cm}|x{.45cm}|x{.4cm}|x{.45cm}|}
\hline
 $0$ &$1$ &$...$ &$1$\\ \hline
 $-1$ &$0$ &$...$ &$0$\\ \hline
 $...$ &$...$ &$...$ &\\ \hline
 $-1$ &$0$ &$...$ &$0$\\ \hline
\end{tabular} 
\setlength\tabcolsep{6pt} 
}
\end{minipage}
\begin{minipage}[c]{.55\linewidth}
\includegraphics[width=1\textwidth]{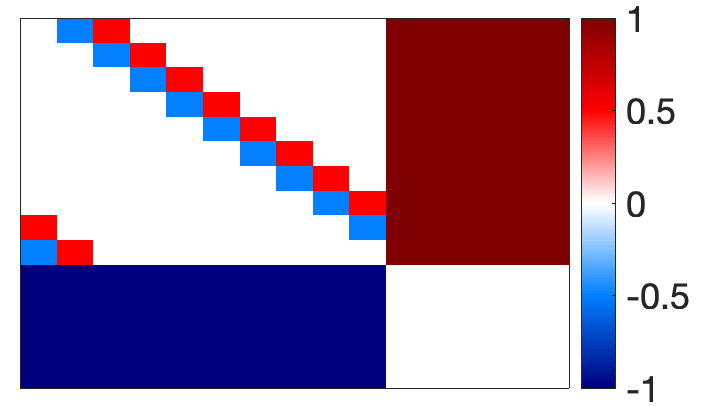}
\end{minipage}
\end{center}

\caption{$R^{\textup{\;eRPS\;}}$ when $k=1$ (left) and $k\ge2$ (right). 
}
\label{tab:eRPS1}
\end{table}

\begin{table*}[t]
\begin{center} 
\begin{tabular}{|c||c|c|c|c|x{.8cm}|c|c|x{.8cm}|x{.8cm}|c|}
\hline
 $\mAa \setminus \mAb$ &$0$ &$1$ &$2$ &$3$ &$...$ &$k - 2$ &$k - 1$ &$k$ &$...$ &$|\mAb| - 1$\\ \hline \hline
$0$ &$0$ &$- \frac{c}{\p_{0} \q_{1}}$ &$\frac{c}{\p_{0} \q_{2}}$ &$0$ &$...$ &$0$ &$0$ &$1$ &$...$ &$1$\\ \hline
$1$ &$0$ &$0$ &$- \frac{c}{\p_{1} \q_{2}}$ &$\frac{c}{\p_{1} \q_{3}}$ &$...$ &$0$ &$0$ &$1$ &$...$ &$1$\\ \hline
$2$ &$0$ &$0$ &$0$ &$- \frac{c}{\p_{2} \q_{3}}$ &$...$ &$0$ &$0$ &$1$ &$...$ &$1$\\ \hline
$3$ &$0$ &$0$ &$0$ &$0$ &$...$ &$0$ &$0$ &$1$ &$...$ &$1$\\ \hline
$...$ &$...$ &$...$ &$...$ &$...$ &$...$ &$...$ &$...$ &$...$ &$...$ &$...$\\ \hline
$k - 2$ &$\frac{c}{\p_{k-2} \q_{0}}$ &$0$ &$0$ &$0$ &$...$ &$0$ &$- \frac{c}{\p_{k-2} \q_{k-1}}$ &$1$ &$...$ &$1$\\ \hline
$k - 1$ &$- \frac{c}{\p_{k-1} \q_{0}}$ &$\frac{c}{\p_{k-1} \q_{1}}$ &$0$ &$0$ &$...$ &$0$ &$0$ &$1$ &$...$ &$1$\\ \hline
$k$ &$-1$ &$-1$ &$-1$ &$-1$ &$...$ &$-1$ &$-1$ &$0$ &$...$ &$0$\\ \hline
$...$ &$...$ &$...$ &$...$ &$...$ &$...$ &$...$ &$...$ &$...$ &$...$ &$...$\\ \hline
$|\mAa| - 1$ &$-1$ &$-1$ &$-1$ &$-1$ &$...$ &$-1$ &$-1$ &$0$ &$...$ &$0$\\ \hline
\end{tabular} \end{center}
\caption{The $R^{\textup{\;eRPS\;}}$ game when $k\ge 2$, i.e. $(\p,\q)$ is a mixed strategy} 
\label{tab:eRPS}
\end{table*}


\begin{lem} \label{lem:erps}
Given any $\left(\p, \q\right)$ with equal support sizes,
the Extended Rock-Paper-Scissors Game $R^{\textup{\;eRPS\;}(\p,\q)}$  has $(\p,\q)$ as the unique Nash equilibrium, and its game value is $0$.
\end{lem}

Note that applying any positive affine transformation to the reward matrix preserves the set of Nash equilibria of the game~\citep*{tewoldegame}. Therefore, if we want the game $R$ to be bounded between $[-b,b]$ for $b>0$, we can simply scale $R^{\textup{\;eRPS\;}}$ by $b$. More generally, for each $\iota>0$ and $v\in\mathbb{R}$, the game $\iota R^{\textup{\;eRPS\;}} + v$ has entries in $\left[v - \iota, v + \iota\right]$ and $\left(\p, \q\right)$ as the unique Nash equilibrium with value $v$. 

There exist other constructions of games that have some $(\p,\q)$ as the unique NE; see \citet{discrete1950hf, nagarajan2020chaos}. Nevertheless, our eRPS construction is simple and intuitive, generalizing the well-known rock paper scissors game. The eRPS game matrix also possesses certain cyclic symmetry and naturally has game value $0$. As we will soon see, the eRPS game is also used in our game modification algorithm. The proof of Lemma~\ref{lem:erps} in Appendix~\ref{app:erps} for eRPS showcases an application of the sufficiency of the SIISOW and INV conditions for NE uniqueness.



\subsection{An Efficient Algorithm for Game Modification in Normal Form Games} \label{sec:lp}

We now turn to the main result of this section: We describe an efficient algorithm to approximately solve Game Modification in normal-form games and provide guarantees on its correctness. In particular, we relax the invertibility constraints so that the remaining constraints are linear and perturb the solution in a way that maintains the feasibility of the linear constraints while making the perturbed solution satisfy invertibility with probability 1.  

Thanks to Lemma~\ref{lem:unique}, the requirement of $R$ having $(\p,\q)$ as the unique NE can be fulfilled by the equivalent SIISOW and the INV conditions, as done in reformulation in Proposition~\ref{thm:gmnf}.
If we ignore the INV condition therein for a moment and tighten the strict inequalities, we obtain an optimization problem with linear constraints: 
\begin{subequations} 
\label{eq:GMLP}
\begin{align}
\displaystyle\min_{R,v} & \;\ell\left(R, R^\circ\right)
\\ \st\ & R_{\I\bullet} \q = v {\bm 1}_{\left| \I \right|} \label{eq:GMLP_SII1}
\\ & \p^\top R_{\bullet\J} = v {\bm 1}^\top_{\left| \J \right|} \label{eq:GMLP_SII2}
\\ & R_{\mAa\setminus\I \bullet} \q \leq \left(v - \iota\right) {\bm 1}_{\left| \mAa\setminus\I \right|} \label{eq:GMLP_SOW1}
\\ & \p^\top R_{\bullet \mAb\setminus\J} \geq \left(v + \iota\right) {\bm 1}^\top_{\left| \mAb\setminus\J \right|} \label{eq:GMLP_SOW2}
\\ & \underline{v} \leq v \leq \overline{v} \label{eq:GMLP_V}
\\ & -b+\lambda \leq R_{i j} \leq b-\lambda, \forall \left(i, j\right) \in \mA. \label{eq:GMLP_B}
\end{align} 
\end{subequations}

In~\eqref{eq:GMLP}, the first four constraints~\eqref{eq:GMLP_SII1}--\eqref{eq:GMLP_SOW2} encode the SIISOW condition. Notice we introduced a small SIISOW margin parameter $\iota>0$ in~\eqref{eq:GMLP_SOW1} and \eqref{eq:GMLP_SOW2}, tightening the strict inequalities in Proposition~\ref{thm:gmnf}. Doing so ensures that the feasible set of the problem~\eqref{eq:GMLP} is closed. A margin $\lambda$ is also added to the reward bound~\eqref{eq:GMLP_B} for reasons that would become clear momentarily.

One can solve the linearly constrained program~\eqref{eq:GMLP} for a solution $R$. To ensure $R$ has a unique NE, it remains to satisfy the INV condition that the matrix 
$\begin{bmatrix} R_{\I \J} & -{\bm 1}_{|\I|} \\ {\bm 1}^\top_{|\J|} & 0 \end{bmatrix}$ must be invertible.
However, enforcing INV directly by constraining the smallest singular value of the matrix leads to a nonlinear, nonconvex optimization problem that is difficult to solve.

We adopt an alternative approach: we take the solution $R'$ to the program~\eqref{eq:GMLP}---which may not satisfy the INV condition---and add a small special random matrix to $R'$ in such a way that: (i) the resulting matrix $R$ is invertible with probability 1; (ii) $R$ still has $(\p,\q)$ as its unique NE and satisfies the value constraint $v\in\left[\underline{v}, \overline{v}\right]$ in~\eqref{eq:GMLP_V}. Moreover, by introducing a small margin $\lambda$ in the reward bound~\eqref{eq:GMLP_B} and using a sufficiently small perturbation,  we further ensure that the perturbed rewards remain in the original designated range $\left[-b, b\right]$. Specifically, the matrix we add is $\varepsilon R^{\textup{\;eRPS\;}}$, where $\varepsilon$ is a random number  in $\left[-\lambda,\lambda\right]$ and $R^{\textup{\;eRPS\;}}$ the Extended Rock-Paper-Scissors game matrix, which has entries in $\left[-1,1\right]$ and game value $0$. 

Combining the above ingredients, we have the complete procedure, Relax And Perturb (RAP), which is presented in Algorithm~\ref{algo:RAP}. RAP approximately solves the Game Modification problem, provably satisfying the constraints with probability~1  and achieving a near minimal cost $\ell\left(R, R^\circ\right)$ as long as the random perturbation is small (Theorem~\ref{thm:perturb}).

\begin{algorithm}[htb]
\caption{Relax And Perturb (RAP)}
\label{algo:RAP}
\textbf{Input}: original game $R^\circ$,
cost function $\ell$,
target policy $\left(\p, \q\right)$, 
target value range $\left[\underline{v}, \overline{v}\right]$,
reward bound $b \in \mathbb{R}^{+} \cup \left\{\infty\right\}$. \\
\textbf{Parameters}: margins $\iota \in \mathbb{R}^{+}$ and $\lambda \in \mathbb{R}^{+}$. \\
\textbf{Output}: modified game $R$. \medskip

\begin{algorithmic}[1] 
\STATE Solve the problem~\eqref{eq:GMLP}. Call the solution $R'$. 
\STATE Sample $\varepsilon \sim \mathrm{uniform}[-\lambda, \lambda]$ 
\STATE Return $R = R'+\varepsilon R^{\textup{\;eRPS\;}\left(\p,\q\right)}$. 
\end{algorithmic}
\end{algorithm}





\begin{rmk}
If the solution of problem~\eqref{eq:GMLP} $R'$ already satisfies the INV condition, that is $\begin{bmatrix} R'_{\I \J} & -{\bm 1}_{|\I|} \\ {\bm 1}^\top_{|\J|} & 0 \end{bmatrix}$ is invertible, then no perturbation is needed. In addition, the perturbation can also be put in a loop while the INV condition is not satisfied, although the perturbed solution satisfies the INV condition with probability one in theory.
\end{rmk}

When the cost function $\ell$ is convex, the problem \eqref{eq:GMLP} is a convex program with linear constraints, for which efficient solvers exist~\citep{wright2006numerical}. The program \eqref{eq:GMLP} is further reduced to a linear program when $\ell$ is piecewise linear, as shown in the following examples.

\begin{eg} [One-Time Cost] \label{eg:lone} 
One can measure the cost of modifying the game from $R^\circ$ to $R$ by the $L^{1}$ distance
\begin{equation*}
\ell\left(R, R^{\circ}\right) = \left\|R - R^{\circ}\right\|_{1}
=\sum_{i \in \mAa, j \in \mAb} \left| R_{ij} - R^{\circ}_{ij} \right|.
\end{equation*}
\end{eg}
This may be interpreted as the one-time cost to the game modifier to take the trouble to change entries in a published payoff matrix.

\begin{eg} [Forever Cost] \label{eg:probweight} 
In some applications, the game modifier incurs a different kind of cost: if the instantiated joint play is row $i$ and column $j$, it costs the game modifier to transfer ``extra money'' $R_{ij} - R^\circ_{ij}$ from the column player to the row player in that round.
For example, this transfer may cost the modifier $|R^\circ_{ij}-R_{ij}|$.
But changes to other payoff matrix entries that are not played do not incur any cost.
Imagine the two players use any reasonable learning algorithms, such as no-regret learning, to play the same modified game forever.
The asymptotic behavior, when game modification is successful and the two players discover the new unique Nash $(\p,\q)$, is that the two will play the game with time average converging to $(\p,\q)$.
The per-round forever cost that the game modifier aims to minimize is then
\begin{equation} \begin{aligned}
\ell\left(R, R^{\circ}\right) &= \displaystyle\sum_{i \in \mAa, j \in \mAb} \p_i \q_j \left| R_{ij} - R^{\circ}_{ij} \right|.
\end{aligned} \end{equation}
An interesting special case is when the target NE $(\p,\q)$ is pure on action pair $(i^*, j^*)$.
Whenever possible, the game modifier should keep the payoff at the NE unchanged:
$$R_{i^*, j^*} = R^\circ_{i^*, j^*}$$
but make other entries in row $i^*$ larger than $R^\circ_{i^*, j^*}$ and in column $j^*$ smaller than $R^\circ_{i^*, j^*}$ to make $(i^*, j^*)$ the unique new NE.
This game modification incurs zero forever cost.
\end{eg}

\subsection{Performance Guarantees for RAP}

Below, we show that the RAP Algorithm has the desired feasibility and near-optimality properties with respect to the original Game Modification problem~\eqref{eq:GMO} in Proposition~\ref{thm:gmnf}. Let $C^\star$ denote the optimal value of \eqref{eq:GMO}. We say that the cost function $\ell$ is $L$-Lipschitz if  $\left| \ell\left(X, R^{\circ}\right) - \ell\left(Y, R^{\circ}\right) \right| \leq L \left\|X - Y \right\|_{1}, \forall X, Y.$

\begin{thm} [Feasibility and Optimality of RAP Algorithm] \label{thm:perturb} 
Suppose that the parameters $\iota, \lambda$ of Algorithm~\ref{algo:RAP} satisfy $$(-b+\lambda+\iota, b-\lambda-\iota)\cap[-\underline{v},\overline{v}] \neq \emptyset$$
and let $R\left(\iota, \lambda\right) = R' + \varepsilon R^{\textup{\;eRPS\;}}$ be the output of the Algorithm~\ref{algo:RAP} with margin parameters $\iota, \lambda$. The following hold.

\begin{enumerate}[itemsep=.5pt, topsep=.5pt]
\item (\textbf{Existence}) The solution $R'$ to program~\eqref{eq:GMLP} exists.

\item (\textbf{Feasibility}) With probability 1, $R\left(\iota, \lambda\right)$ is feasible for the original Game Modification problem~\eqref{eq:GMO}.

\item (\textbf{Optimality}) If in addition the cost  $\ell$ is $L$-Lipschitz with $L<\infty$, then $R\left(\iota, \lambda\right)$ is asymptotically optimal:
$$\lim_{\max\left\{\iota, \lambda\right\} \to  0} \ell\left(R\left(\iota, \lambda\right), R^\circ\right) = C^\star.$$ 
\item (\textbf{Optimality Gap}) If $\ell$ is piecewise linear (e.g., $L^1$ cost), then the optimality gap is at most linear in $(\iota,\lambda)$: 
$$\ell\left(R\left(\iota, \lambda\right), R^\circ\right) = C^\star + O(\max\left\{\iota, \lambda\right\}).$$
\end{enumerate}
\end{thm}

In the result above, existence follows from Theorem~\ref{thm:feasible}. Feasibility holds because the matrix sum 
\begin{equation*}
\begin{bmatrix} R'_{\I\J} & -{\bm 1}_{\left| \I \right|} \\ {\bm 1}_{\left| \J \right|}^\top & 0 \end{bmatrix}
+ \varepsilon
\begin{bmatrix} R^{eRPS}_{\I\J} & -{\bm 1}_{\left| \I \right|} \\ {\bm 1}_{\left| \J \right|}^\top & 0 \end{bmatrix}
\end{equation*}
is invertible with probability $1$, as $\varepsilon$ is a continuous random variable and the second matrix above is invertible. 

To prove optimality, we take a feasible solution $R^{\left(\varepsilon\right)}$  to the original game modification problem \eqref{eq:GMO} with a cost at most $C^\star+\varepsilon$, and then slightly and carefully modify its entries to get a new solution $R'^{\left(\varepsilon\right)}$ for which (i) the reward bound~\eqref{eq:GMLP_B} with $\lambda$ margin is satisfied, (ii) the SIISOW properties~\eqref{eq:GMLP_SII1}--\eqref{eq:GMLP_SOW2} are preserved, and (iii) the game value is the same. The costs of $R'^{\left(\varepsilon\right)}$ and $R^{\left(\varepsilon\right)}$ are close thanks to the Lipschitz property of the cost. In particular, the difference $\ell(R'^{(\varepsilon)} ,R^{\circ}) -\ell(R^{\left(\varepsilon\right)}, R^{\circ})$, and in turn the optimality gap, vanish when the margin parameters $\iota, \lambda$ go to zero. 

Part 4 of the theorem further shows that the optimality gap vanishes at a linear rate in $(\iota, \lambda)$ under piecewise linear cost $\ell$. In this case \eqref{eq:GMLP} is a linear program with a full rank constraint matrix, and we can control the optimality gap using techniques from sensitivity analysis of linear programs ~\citep{bertsimas1997introduction, jansen1997sensitivity}.

\section{Markov Games Modification}
\label{sec:markovgame}
In this section, we generalize to Markov games. We install a possibly stochastic policy as the unique Markov perfect equilibrium by installing a unique Nash equilibrium in every stage game defined by the Q functions. 

\subsection{Preliminaries}

A finite-horizon two-player zero-sum Markov game can be described by a pair $(P, R)$, given the finite state space $\mathcal{S}$, the finite joint action space $\mA = \mAa \times \mAb$, and horizon $H$. Here $P = \big\{P_{h} : \mathcal{S} \times \mathcal{S} \to  [0, 1]^{| \mAa | \times | \mAb |}\big\}_{h=1}^{H}$ is the transition probabilities, $P_{0} : \mathcal{S} \to  [0, 1]$ the initial state distribution, and $R = \big\{R_{h} : \mathcal{S} \to [-b, b]^{| \mAa| \times | \mAb|}\big\}_{h=1}^{H}$ the mean reward function. For each  $h \in \left[H\right], s \in \mathcal{S}$, we treat $R_{h}(s)$ as an $\left| \mAa \right| \times \left| \mAb \right|$ matrix, where $\left[R_{h}(s)\right]_{ij}$ is the reward for the joint action profile $(i,j)\in \mAa \times \mAb$. Similarly, the transition probabilities are given by an $| \mAa | \times | \mAb |$ matrix  $P_{h}\left(s' | s\right)$, where $[P_{h}(s' | s)]_{ij}$ is the probability of transitioning from state $s \in \mathcal{S}$ in period $h \in \left[H\right]$ to state $s' \in \mathcal{S}$ when the joint action $(i,j)$ is used. 
The above matrix representations are chosen to follow the convention used in the last section for normal-form matrix games.

A Markovian policy $(\p, \q)$ is a pair of policies for the two players: $\p = \{\p_{h} : \mathcal{S} \to  \Delta_{\mAa} \}_{h=1}^{H}$ and $\q = \{\q_{h} : \mathcal{S} \to  \Delta_{\mAb}\}_{h=1}^{H}$. Here
$\p_{h}(s)$ and $\q_{h}(s)$ are probability vectors; in period $h \in [H]$, state $s \in \mathcal{S}$, $[\p_{h}(s)]_{i}$ specifies the probability that player 1 takes action $i \in \mAa$; 
similarly for $[\q_{h}(s)]_{j}$.


A zero-sum Markov game has at least one Markov perfect equilibrium and a unique Nash value. The action-value or Q function of the MPE, denoted by $Q^\star$, satisfies the following Bellman equations: for each $h \in \left[H\right], s \in \mathcal{S}, \left(i, j\right) \in \mA$,
\begin{align}
& Q^\star_{h}\left(s, \left(i, j\right)\right) \coloneqq R_{h}\left(s, \left(i, j\right)\right) + 
 \label{eq:bellman}\\ 
& \hspace{1em} \displaystyle\sum_{s' \in \mathcal{S}} P_{h}\left(s' | s, \left(i, j\right)\right) \displaystyle\max_{p' \in \Delta_{\mAa}} \displaystyle\min_{q' \in \Delta_{\mAb}} Q^\star_{h+1}\left(s', \left(p', q'\right)\right), \nonumber
\end{align}
where for a possibly stochastic strategy profile $\left(p', q'\right)\in \Delta_{\mAa} \times \Delta_{\mAb}$, we define
\begin{equation} \begin{aligned}
Q^\star_{h}\left(s, \left(p', q'\right)\right) \coloneqq \displaystyle\sum_{i \in \mAa, j \in \mAb} p'_{i} q'_{j} Q^\star_{h}\left(s, \left(i, j\right)\right).
\end{aligned} \end{equation}
We use the convention that $Q^\star_{H+1}\left(s, \left(i, j\right)\right) = 0,\forall s,i,j$.

Under an MPE policy, the stage game of the Markov game in each period $h \in \left[H\right]$ and state $s \in \mathcal{S}$ is a normal form game with payoff matrix $\mQ_{h}\left(s\right)$, whose $(i,j)$ entry is
\begin{equation} \begin{aligned}
[\mQ_{h}\left(s\right)]_{ij} \coloneqq Q^\star_{h}\left(s, \left(i,j\right)\right)
\label{eq:Q_matrix}
\end{aligned} \end{equation}
and corresponds to the payoff under the action profile $\left(i, j\right) \in \mA$.
Consequently, an MPE can be defined recursively as the Nash equilibrium for every stage game.

\begin{df} [Markov Perfect Equilibrium] \label{df:mpe} 
A Markov perfect equilibrium  policy $\left(\p, \q\right)$ is a policy that satisfies, for every $h \in \left[H\right], s \in \mS$,
\begin{equation*} \begin{aligned}
\left(\p_{h}(s), \q_{h}(s)\right) \text{is a Nash equilibrium of }\mQ_{h}(s),
\end{aligned} \end{equation*}
where $\mQ_h(s)$ is defined by equations~\eqref{eq:bellman}--\eqref{eq:Q_matrix}.
\end{df}

We remark that an alternative approach to studying the equilibria of Markov games is by converting it to a single, big normal-form game and considering the NEs of the latter. An NE defined in this way is, in general, not Markov perfect---it requires coordination and commitment to policies in stage games that are not visited along equilibrium paths. Such policies are often not realistic. Moreover, it is computationally intractable to manipulate such a big normal-form game. Therefore, we focus on MPEs and make use of their recursive characterization through the Bellman equations.

\subsection{Reformulation and Feasibility of Markov Game Modification}

A two-player zero-sum Markov game has a unique MPE if and only if every stage game $\mQ_h(s)$ has a unique NE. Our results on the uniqueness of NE for normal form games (Lemma~\ref{lem:unique}) apply to each stage game of the Markov game. Combining these two observations and the Bellman equations for $\mQ_h(s)$'s, we can write the Game Modification problem in Definition \ref{def:GM} for Markov games equivalently as an optimization problem similar to~\eqref{eq:GMO}, where SIISOW (Condition~\ref{def:gap}), INV (Condition~\ref{df:inv}) and the Bellman equations are imposed as constraints for every stage game. Due to space limit, this optimization problem is provided in the appendix.

We provide a sufficient and necessary condition for the feasibility of the above Game Modification problem for Markov games. Let $\I_h(s)=supp(\p_h(s))$ and $\J_h(s)=supp(\q_h(s))$.

\begin{thm} [Feasibility of Markov Game Modification] \label{cor:gamefeas} 
The Game Modification problem in Definition~\ref{def:GM} for Markov games is feasible if and only if $\left| \I_{h}\left(s\right) \right| = \left| \J_{h}\left(s\right) \right|$ for every $h \in \left[H\right], s \in \mathcal{S}$, and $\left(- H b, H b\right) \cap \left[\underline{v}, \overline{v}\right] \neq \emptyset$.
\end{thm}

The above theorem subsumes Theorem~\ref{thm:feasible} for normal-form games. The sufficient condition above is proved by explicitly constructing a feasible Markov game, recursively using the Extended Rock-Paper-Scissors game. 

\subsection{Efficient Algorithm for Modifying Markov Games} \label{sec:mlp}

To develop an efficient algorithm, we follow a similar strategy as in normal form games: we ignore the INV (invertibility) condition and retain only the linear constraints for the Markov game modification problem, and add small margins $\iota,\lambda$ to the SIISOW and reward bound constraints so that random perturbation can be added later. Doing so leads to a linearly constrained optimization problem, given in~\eqref{eq:MGMLP}, which generalizes the program~\eqref{eq:GMLP} for normal-form games.  

\begin{align}
\min_{R,v,\mQ} & \;\ell\left(R, R^{\circ}\right)
\label{eq:MGMLP} \\ \st & \left[\mQ_{h}\left(s\right)\right]_{\I_{h}\left(s\right)\bullet} \q_{h}\left(s\right) = v_{h}\left(s\right) {\bm 1}_{\left|\I_{h}\left(s\right)\right|}
\notag\\ & \hspace{2em} \forall\; h \in \left[H\right], s \in \mathcal{S} \hspace{3.05cm}\textup{[row SII]}
\notag\\ & \p^\top_{h}\left(s\right) \left[\mQ_{h}\left(s\right)\right]_{\bullet\J_{h}\left(s\right)} = v_{h}\left(s\right) {\bm 1}^\top_{\left|\J_{h}\left(s\right)\right|}
\notag\\ & \hspace{2em} \forall\; h \in \left[H\right], s \in \mathcal{S}
\hspace{2.5cm}\textup{[column SII]}
\notag\\ & \left[\mQ_{h}\left(s\right)\right]_{\mAa \setminus \I_{h}\left(s\right) \bullet} \q_{h}\left(s\right) \leq \left(v_{h}\left(s\right)-\iota\right) {\bm 1}_{\left|\mAa \setminus \I_{h}\left(s\right)\right|}
\notag\\ & \hspace{2em} \forall\; h \in \left[H\right], s \in \mathcal{S} \hspace{2.7cm}\textup{[row SOW]}
\notag\\ & \p^\top_{h}\left(s\right) \left[\mQ_{h}\left(s\right)\right]_{\bullet \mAb \setminus \J_{h}\left(s\right)} \geq \left(v_{h}\left(s\right)+\iota\right) {\bm 1}^\top_{\left|\mAb \setminus \J_{h}\left(s\right)\right|}
\notag\\ & \hspace{2em} \forall\; h \in \left[H\right], s \in \mathcal{S}
\hspace{2.15cm}\textup{[column SOW]}
\notag\\ & \mQ_{h}\left(s\right) = R_{h}\left(s\right) + \displaystyle\sum_{s' \in \mathcal{S}} P_{h}\left(s' | s\right) v_{h+1}\left(s'\right)
\notag\\ & \hspace{2em} \forall\; h \in \left[H - 1\right], s \in \mathcal{S} \hspace{2.3cm}\textup{[Bellman]}
\notag\\ & \mQ_{H}\left(s\right) = R_{H}\left(s\right), \forall\; s \in \mathcal{S}
\notag\\ & \underline{v} \leq \displaystyle\sum_{s \in \mathcal{S}} P_{0}\left(s\right) v_{1}\left(s\right) \leq \overline{v} \hspace{2cm}\textup{[value range]}
\notag\\ & -b+\lambda \leq \left[R_{h}\left(s\right)\right]_{i j} \leq b-\lambda
\notag\\ & \hspace{2em} \forall\; \left(i, j\right) \in \mA, h \in \left[H\right], s \in \mathcal{S} 
\hspace{.5cm}\textup{[reward bound]}
\notag
\end{align}

\begin{rmk}
If there is no value range constraint and the cost $\ell$ is decomposable across the states and periods (e.g., $L^1$ cost), then the program~\eqref{eq:MGMLP} can be broken into $H \left|\mS\right|$ smaller optimization problems, one for each stage game, that can be solved sequentially by backward induction. 
\end{rmk}

We present our algorithm, Relax And Perturb for Markov Games (RAP-MG), in Algorithm~\ref{algo:RAP2}, which adds random perturbation to the reward matrix of every stage game. 

\begin{algorithm}[htb]
\caption{Relax And Perturb for Markov Games (RAP-MG)}
\label{algo:RAP2}
\textbf{Input}: original game $\left(R^\circ, P\right)$,
cost function $\ell$,
target policy $\left(\p, \q\right)$ and value range $\left[\underline{v}, \overline{v}\right]$,
reward bound $b \in \mathbb{R}^{+} \cup \left\{\infty\right\}$. \\
\textbf{Parameters}: margins $\iota \in \mathbb{R}^{+}$ and $\lambda \in \mathbb{R}^{+}$. \\
\textbf{Output}: modified game $\left(R,P\right)$. \medskip

\begin{algorithmic}[1] 
\STATE Solve the problem~\eqref{eq:MGMLP}. Call the solution $R'$.
\FOR {$h \in \left[H\right], s \in \mS$}
    \STATE Sample $\varepsilon \sim \mathrm{uniform}[-\lambda, \lambda]$ 
    \STATE Perturb the reward matrix in stage $\left(h, s\right)$:
    \STATE \hspace{1 em} $R_{h}\left(s\right) = R'_{h}\left(s\right)+\varepsilon R^{\textup{\;eRPS\;}\left(\p_{h}\left(s\right),\q_{h}\left(s\right)\right)}$. 
\ENDFOR
\STATE Return $\left(R,P\right)$.
\end{algorithmic}
\end{algorithm}

In the theorem below we provide feasibility and optimality guarantees for Algorithm~\ref{algo:RAP2}. These results are similar to those normal form games in Theorem~\ref{thm:perturb}, but the proofs are more complicated due to the dependency across the stage games. Let $C^\star$ be the optimal objective value for the original game modification problem in Definition~\ref{def:GM}.

\begin{thm} [Feasibility and Optimality of the RAP-MG Algorithm] \label{cor:perturbmg} 
Let $R\left(\iota, \lambda\right) = R' + \varepsilon R^{\textup{\;eRPS\;}}$ denote the output of Algorithm~\ref{algo:RAP2} with margin parameters $\iota, \lambda$. If
\begin{equation} \label{eq:li}
(-b+\lambda+\iota, b-\lambda-\iota)\cap \left[-\underline{v}/H, \overline{v}/H \right] \neq \emptyset,
\end{equation}
then the following hold.
\begin{enumerate}
\item (\textbf{Existence}) The solution $R'$ to the program~\eqref{eq:MGMLP} exists.
\item (\textbf{Feasibility}) $R\left(\iota, \lambda\right)$ is feasible for the game modification problem in Definition~\ref{def:GM} with probability $1$.
\item (\textbf{Optimality}) If in addition the cost function $\ell$ is $L$-Lipschitz, then
$R\left(\iota, \lambda\right)$ is asymptotically  optimal:
\begin{equation} 
\lim_{\max\left\{\iota, \lambda\right\} \to  0} \ell\left(R\left(\iota, \lambda\right), R^\circ\right) = C^\star,\nonumber
\end{equation}
\item (\textbf{Optimality Gap}) If $\ell$ is piecewise linear, then
\begin{equation} 
\ell\left(R\left(\iota, \lambda\right), R^\circ\right) = C^\star + O(\max\left\{\iota, \lambda\right\}),\nonumber
\end{equation}
\end{enumerate}
\end{thm}

\section{Experiments}
Our code is available at: \url{https://github.com/YoungWu559/game-modification}.

\subsection{Toy Experiments}

We run Algorithm~\ref{algo:RAP} on several small normal-form games such as two-finger Morra and five-action rock-paper-scissors games.

\paragraph{1.}
Given left below is the payoff matrix for the \textbf{simplified Two-finger Morra}  game~\citep{good1965f}, which has a unique NE $(\p,\q) = (\frac{7}{12}, \frac{5}{12})$ and value $-\frac{1}{12}$. On the right we minimally modify the game to keep the same unique NE but make the game fair with a value of $0$.
\[
\text{Original: }
    \begin{pmatrix}
        2 & -3 \\
        -3 & 4
    \end{pmatrix}
\quad 
\text{Modified: }
    \begin{pmatrix}
        2.04 & -2.86 \\
        -2.86 & 4
    \end{pmatrix}
\]
We provide another example of game modification for the classic Two-finger Morra game in the Appendix~\ref{app:exp}.
\paragraph{2.}
The \textbf{Rock-Paper-Scissors-Fire-Water} game, given on the left below, is a generalization of the Rock-Paper-Scissor game to five actions~\citep{tagiew2009hypotheses}. 
The unique NE is $\p=\q=(\frac{1}{9}, \frac{1}{9},\frac{1}{9},\frac{1}{3}, \frac{1}{3})$ and has value $0$. We desire the NE to be simpler for humans, so we redesign the game to have a uniformly mixed NE $\p=\q = (\frac{1}{5}, \frac{1}{5},\frac{1}{5},\frac{1}{5},\frac{1}{5})$. The resultant game is given below.

Note that an alternative 5-action game, Rock-Paper-Scissors-Spock-Lizard, also has the desired NE (more details are provided in the Appendix~\ref{app:exp}). However, our modification has a lower modification cost $4$, compared to the cost $8$ for using the alternative game.
\begin{align*}
    \text{Original} \qquad\qquad\;  & \qquad\qquad\;  \text{Modified} \\
    \begin{pmatrix}
    0 & -1 & 1 & -1 & 1 \\
    1 & 0 & -1 & -1 & 1 \\
    -1 & 1 & 0 & -1 & 1 \\
    1 & 1 & 1 & 0 & -1 \\
    -1 & -1 & -1 & 1 & 0 
    \end{pmatrix}
    &
 \begin{pmatrix}
    0 & -1 & 1 & -1 & 1 \\
    1 & 0 & -1 & -1 & 1 \\
    -1 & 1 & 0 & -1 & 1 \\
    1 & 1 & 1 & 0 & -3  \\
    -1 & -1 & -1 & 3 & 0 
    \end{pmatrix}
\end{align*}

\subsection{Approximation}

Theorem~\ref{thm:perturb} shows that Algorithm~\ref{algo:RAP} approaches the optimal cost $C^*$ as a linear function in $\max\{\iota, \lambda \}$ in the worst case. To see how fast the convergence happens in practice, we tested RAP on a fixed Game Modification instance with varying choices of $\iota$ and $\lambda$. In particular, we considered $(p,q) = ((.47,.53,0,0)^{\top}, (.42,.58,0,0)^{\top})$,
\begin{equation}
    R^{\circ} = \begin{pmatrix}
        -0.33 & -0.03 & 0.68 & -0.04 \\
         0.16 & -0.43 & 0.94 & -0.45 \\
         0.02 & 0.85 & -0.28 & -0.98 \\
        -0.57 & 0.3  & -0.12 & -0.17 \\
    \end{pmatrix},
\end{equation}
and no reward bound or value constraints. We considered $\iota$ and $\lambda$ of the form $10^{-i}$ for $i \in \{0, \ldots, 15\}$. However, convergence happened by $10^{-4}$ for both parameters and so we only report for parameters down to $10^{-4}$. 

\begin{figure}
    \centering
    \includegraphics[scale=.40]{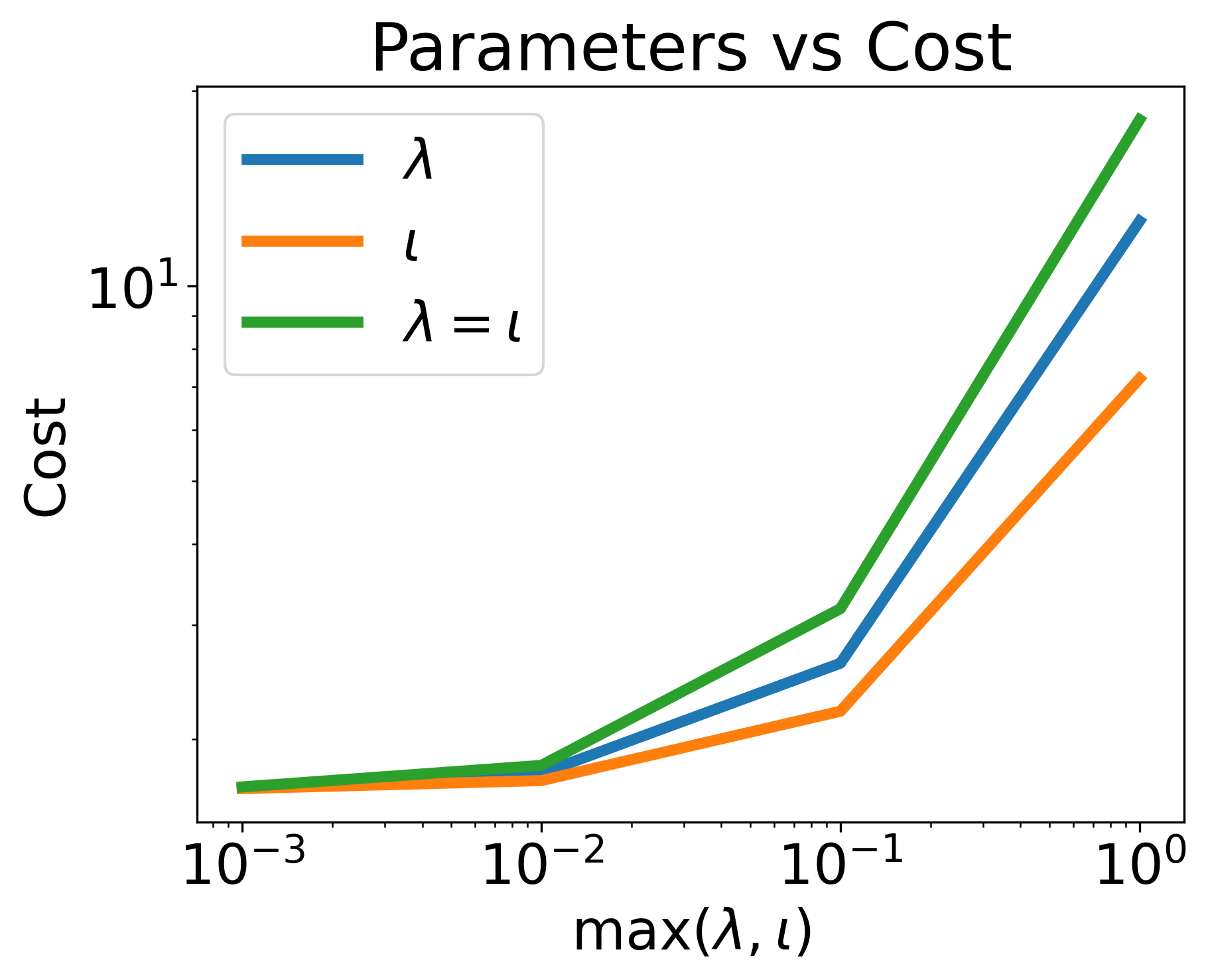}
    \caption{Convergence to Optimal Cost}
    \label{fig:approx}
\end{figure}

To further explore which parameter had the largest effect on the cost, we ran RAP under three different configurations. The result is Figure~\ref{fig:approx}. First, we fixed $\iota = 10^{-5}$ and varied $\lambda$ to construct the $\lambda$ curve. Second, we fixed $\lambda = 10^{-5}$ and varied $\iota$ to construct the $\iota$ curve. Lastly, we varied both equally, i.e. considered $(\iota,\lambda) = (10^{-i},10^{-i})$, to construct the $\lambda = \iota$ curve. 

We observe that in all three cases, convergence happened even faster than the linear rate promised by Theorem~\ref{thm:perturb}. In addition, we see that $\lambda$ was generally the bottleneck for convergence with the $\lambda$ curve being very close to the $\lambda = \iota$ curve. In contrast, $\iota$ had less of an impact on convergence. We ran the same experiment on other, uniform-randomly generated instances and noticed a general trend of $\lambda$ being the dominant factor. 

\subsection{Scale Benchmarks}

\begin{figure}[t] 
\begin{minipage}[t]{.49\linewidth}
\includegraphics[width=\linewidth]{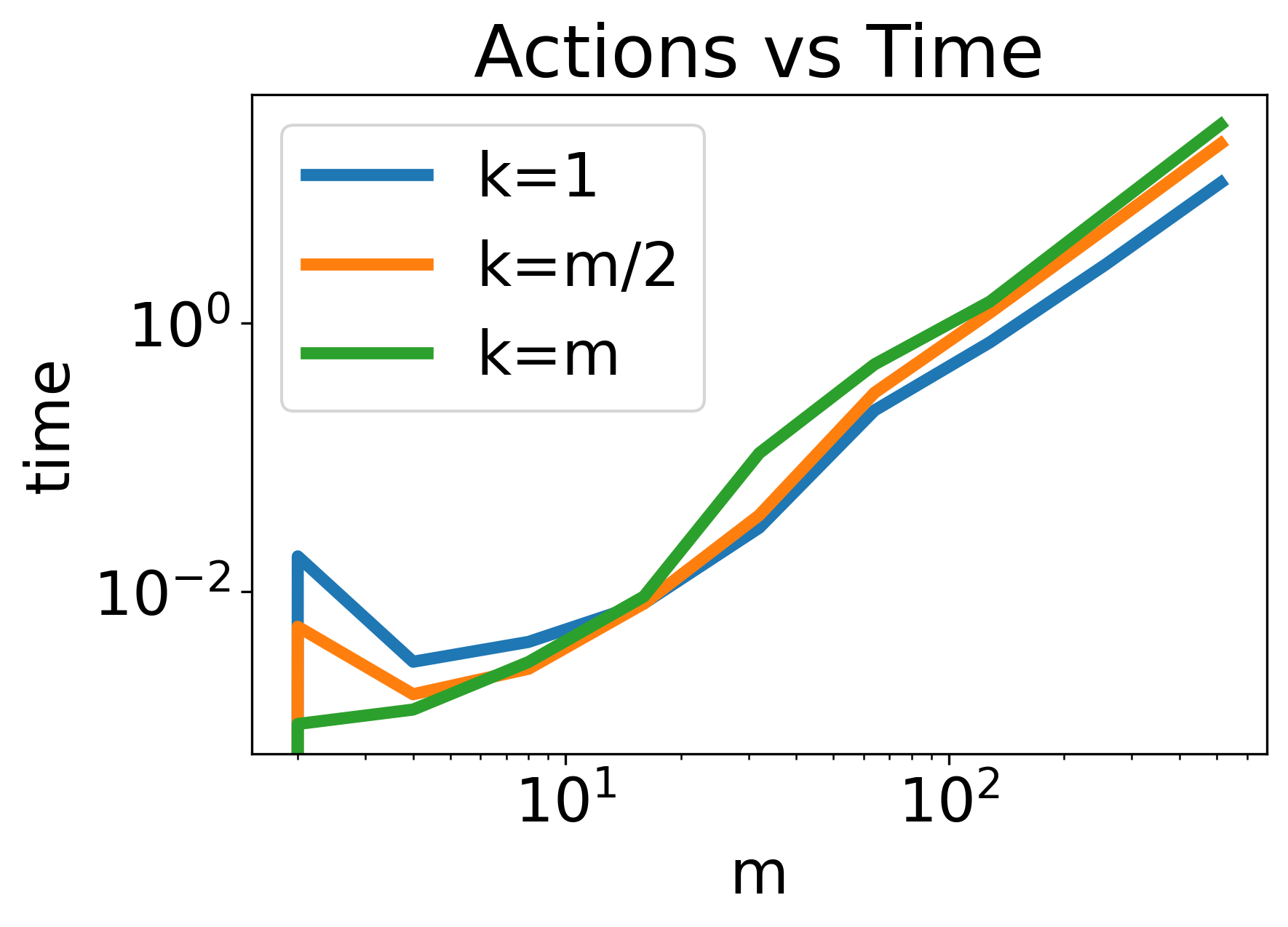}
\end{minipage}
\begin{minipage}[b]{.49\linewidth}
\includegraphics[width=\linewidth]{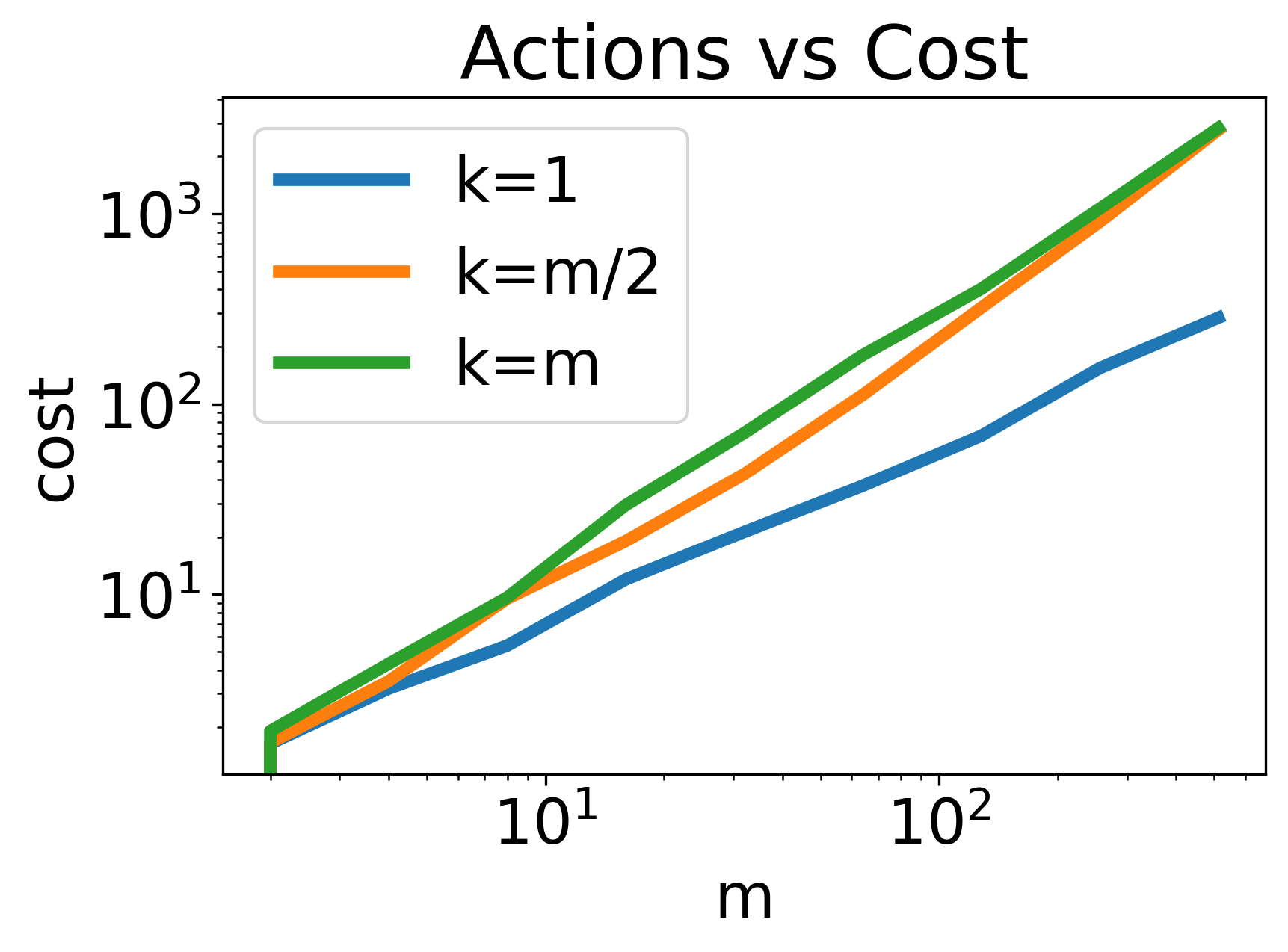}
\end{minipage}
\caption{Scale Benchmark for Number of Actions} \label{fig:time1}
\end{figure}

\begin{figure}[t]
\begin{minipage}[t]{.49\linewidth}
\includegraphics[width=\linewidth]{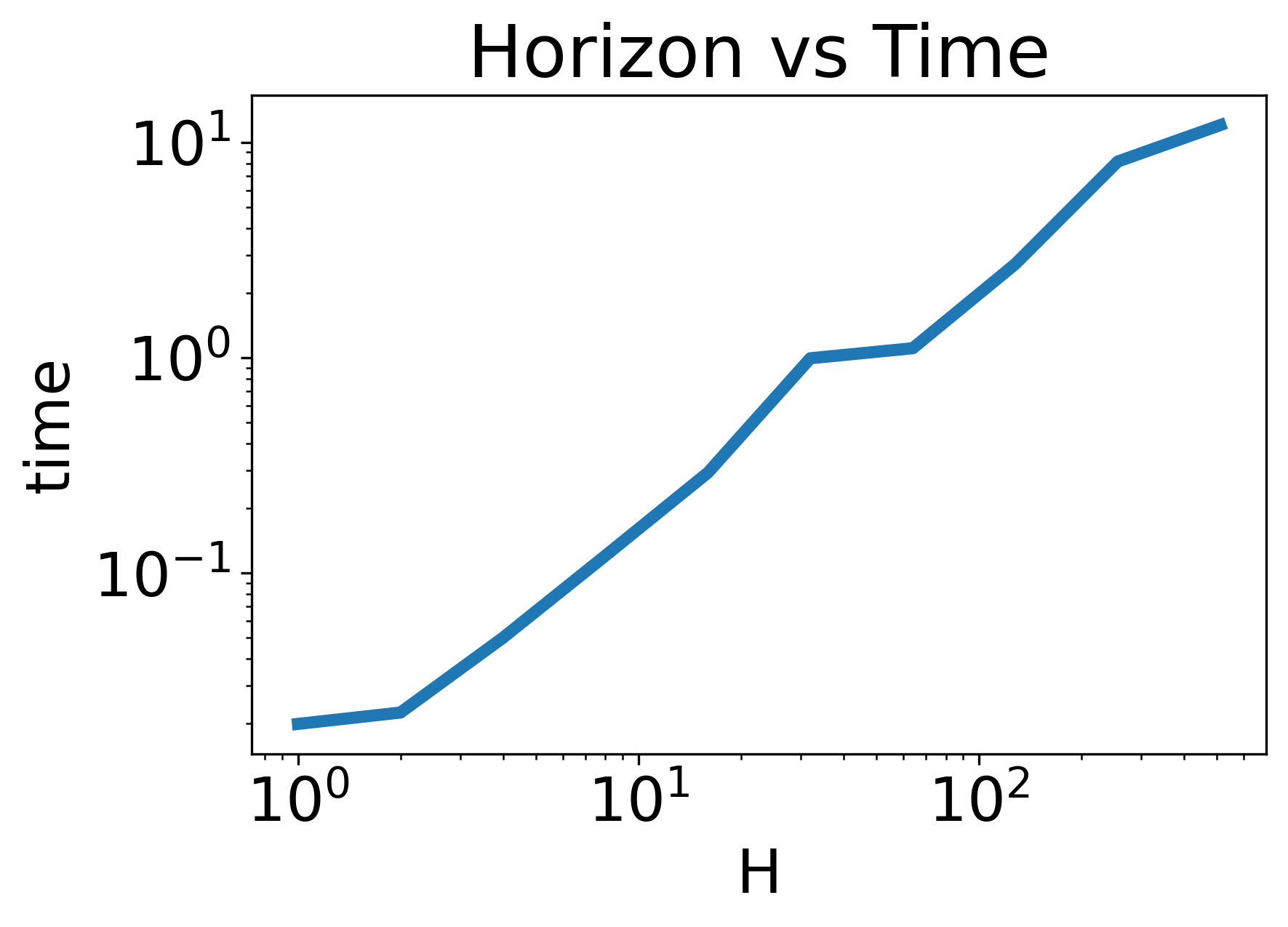}
\end{minipage}
\begin{minipage}[b]{.49\linewidth}
\includegraphics[width=\linewidth]{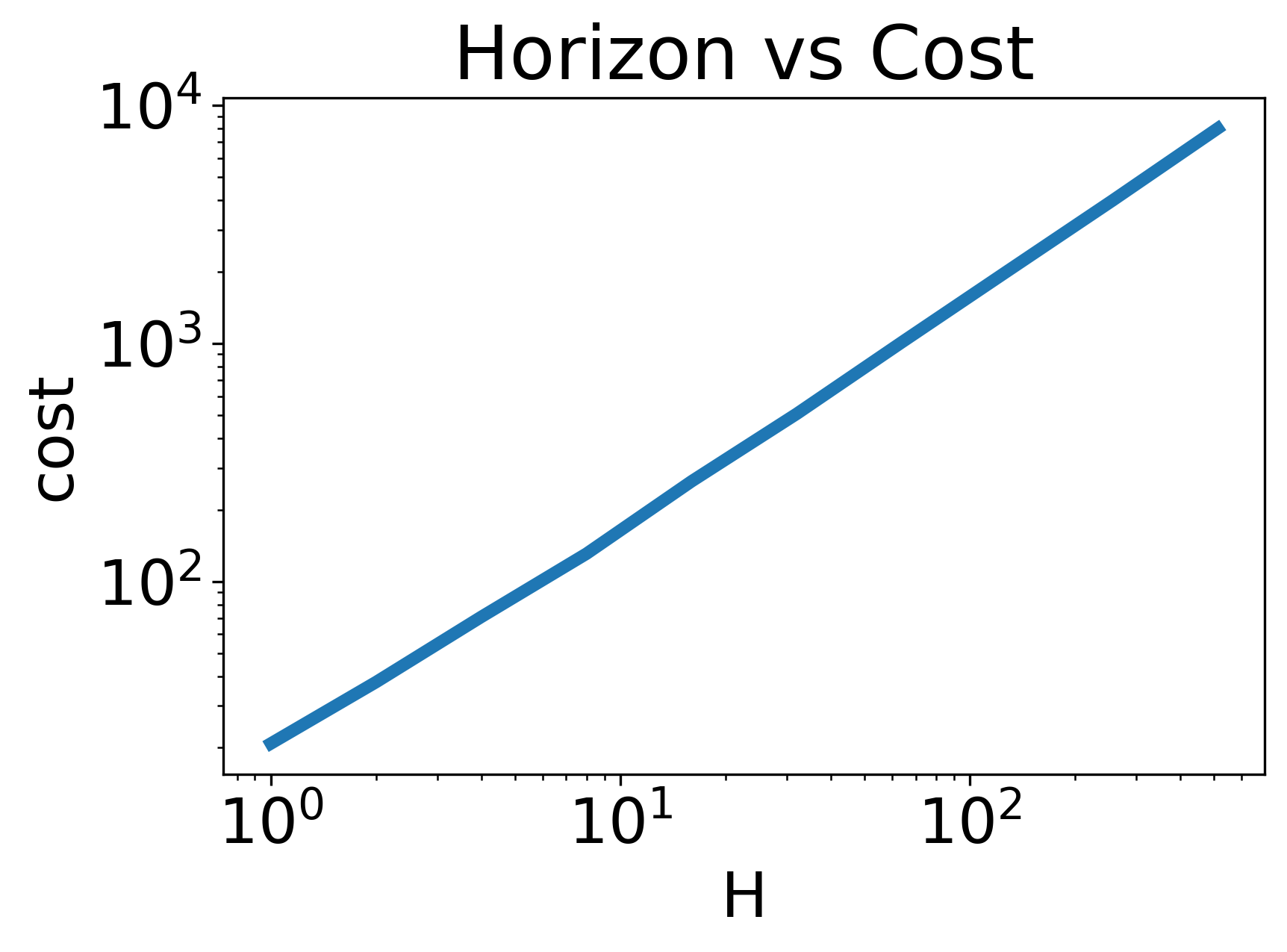}
\end{minipage}
\caption{Scale Benchmark for Number of Periods} \label{fig:time2}
\end{figure}


We run Algorithm~\ref{algo:RAP} and Algorithm~\ref{algo:RAP2} on several games to illustrate the efficacy of our techniques. 
We know our algorithm succeeds by checking that $(\p,\q)$ satisfies the SIISOW and INV conditions for $R^{\dagger}$. 

We first show how our methods scale with the number of actions. For each $m \in \{2,4,8, \ldots, 512\}$ we generate $N = 5$ random matrices $R^{\circ} \sim \mathrm{uniform}[-1,1]^{m \times m}$. For each matrix, we also generate $3$ random $(\p,\q) \sim \text{Dirichlet}(1,\ldots,1)$ with support size (i) $k = 1$, (ii) $k = m/2$, and (iii) $k = m$ (full support). We run Algorithm~\ref{algo:RAP} on each instance and report the worst running time (in seconds) and the worst cost encountered for each $m$ in Figures~\ref{fig:time1}. We see that the solving time grows linearly in the log, so the runtime is polynomial in the actions. Using the Gurobi LP solver, even on a laptop computer, the algorithm handles millions of variables ($512^2$) in roughly $10$ seconds. The $L^1$ costs also appear to grow linearly, though with different slopes. 

Next, we show how our methods scale with the horizon. We consider Markov games with $S = 10$, $A = 2$, random transitions and random reward matrices. Formally, for each $H \in \{1, 2, 4, \ldots, 512\}$, we generate $N = 5$ random Markov games and corresponding target NE pairs with full support. For any fixed $H$, we generate $R_h(s) \in \mathrm{uniform}[-1,1]^{2 \times 2}$ for each $h$ and $s$, and choose $P_h(s,a) \sim \text{Dirichlet}(1,\ldots 1)$ for each $(h,s,a)$. We run Algorithm~\ref{algo:RAP2} on each instance and report the worst running time and cost encountered for each $H$ in Figures~\ref{fig:time2}. We observe the solutions are correct, and again, the algorithm is efficient.

\section{Concluding Remarks}
\label{sec:conclusion}

Our work points to several future directions: (i) It is interesting to study Markov game modification problems where the transition probabilities can also be changed and generalize to general-sum, multi-agent games with other equilibrium concepts. (ii)  In many games, the rewards are constrained to take discrete values (e.g., $-1, 0, 1$). The feasibility and tractability of such constrained game modification problems require further investigation. (iii) It is non-trivial to extend the problem when the attacker's target is an infinite set of policies, for example, when the attacker only cares about the support of the target policies and not specific mixing probabilities or when the attacker only cares about the target policies of one of the players. (iv) Extending our results to data poisoning problems, where the players learn the true game from observational data, leads to interesting theoretical and algorithmic questions.



\section*{Acknowledgments}
Xie was supported in part by National Science Foundation Awards CNS-1955997 and EPCN-2339794. Zhu was supported in part by NSF grants 1836978, 2023239, 2202457, 2331669, ARO MURI W911NF2110317, and AF CoE FA9550-18-1-0166. Chen is supported in part by NSF grant CCF-2233152. We would like to thank Lijun Ding for inspiring discussion. We would like to thank Joy Cheng for implementing the code for the RAP algorithm.

\section*{Impact Statement}
\emph{This paper presents work whose goal is to advance the field of MARL. Our work is largely theoretical, so we do not see any immediate negative societal impacts.}

\bibliography{icml2024}
\bibliographystyle{icml2024}

\appendix
\onecolumn

\section{Appendix}

In this appendix we provide omitted proofs and additional experiments.

\subsection{Proof of Lemma~\ref{lem:unique}}\label{app: unique}

\begin{proof}

Lemma~\ref{lem:unique} states that the SIISOW and INV conditions are sufficient and necessary for $(\p,\q)$ to be the unique NE of the game $R$. We prove sufficiency and necessity separately.

We exploit the well-established connection between Nash equilibrium and linear program duality. In particular, any $(\pia,\pib)$ that is a Nash equilibrium of $R$ is an optimal solution pair to the following pair of primal-dual linear programs (LPs), and vice versa \citep{dantzig1963linear}.

\begin{df} [Linear Programs for NE] \label{df:lp} 
\begin{equation} 
\begin{aligned}
\textup{(Primal)}\quad \displaystyle\max_{\pia' \in \Delta\mAa, v} \; & v
\\ \st \; & \pia'^\top \mR \ge v {\bm 1}_{|\J|}^\top
\end{aligned} \end{equation}
\begin{equation} 
\begin{aligned}
\textup{(Dual)}\quad \displaystyle\min_{\pib' \in \Delta\mAb, v} \; & v
\\ \st \; & \mR \pib' \le v {\bm 1}_{|\I|}
\end{aligned} \end{equation}
The inequalities are elementwise.
\end{df}
The optimal values of the two linear programs both equal $v^*$, the value of the game.

We emphasize that these LPs are used only for characterizing the properties of the set of Nash equilibria of $R$ and its uniqueness. We do \emph{not} assume that the players must use LP to find an NE: they can use any other solvers and may find any one of the NEs if there are multiple ones. This reflects how NE solvers typically work in practice.

\textbf{Conditions $\Rightarrow$ unique NE:}
We have already argued that $(\p,\q)$ is an NE; see the discussion after the definition of SIISOW.  Suppose $(\r,\s)$ is another NE. We show that it must be the case $\r=\p, \s=\q$.

First of all, it is easy to see that $supp(\r)\subseteq \I, supp(\s)\subseteq \J$.
Suppose there is a violation $\exists i \in supp(\r), i \notin \I$.
By~\eqref{eq:outpworse}, 
$\e_i^\top R \q < \p^\top R \q=v^*$ which leads to $\r^\top R\q < v^*$.
But since $(\r,\s)$ is another NE in a two-player zero-sum game, $(\r,\q)$ is a third NE with $\r^\top R \q=v^*$
, a contradiction.
The case for $\s$ is similar.

Because $(\r,\s)$ is an NE, it satisfies the primal-dual LP in Definition~\ref{df:lp}.
Now with the support constraints, they satisfy the reduced LPs where the vectors and matrices are restricted to the appropriate support:
\begin{eqnarray}
\max_{{\r'}_\I \in \Delta_\I, v} v && \mbox{s.t.} \; {\r'}_\I^\top R_{\I\cdot} \ge v {\bm 1}_{|\J|}^\top \\
\min_{{\s'}_\J \in \Delta_\J, v} v && \mbox{s.t.} \; R_{\cdot\J} {\s'}_\J \le v {\bm 1}_{|\I|}.
\end{eqnarray}
We now show this must mean $\s=\q$.
Consider two cases on the dual restricted LP:

(Case 1) At the solution $(\s,v^*)$, all constraints in $R_{\I\J} \s_\J \le v^*$ are active, i.e. they are equalities $R_{\I\J} \s_\J = v^*$.
Also $\s_\J$ sums to 1. We may write the two as a linear system:
\begin{equation}
\begin{bmatrix} R_{\I\J} & -1 \\ 1 & 0 \end{bmatrix} 
\begin{bmatrix} \s_\J \\ v^* \end{bmatrix} 
= 
\begin{bmatrix} 0 \\ 1 \end{bmatrix}. 
\end{equation}
By the invertability condition, $\s_\J$ has a unique solution and it must equal $\q_\J$ because $\q_\J$ is also a solution to this linear system. The rest of $\s$ and $\q$ are both zeros.  Thus $\s=\q$.

(Case 2)
At least one constraint in $R_{\I\J} \s_\J \le v^*$ is inactive. Then there exists slack variables $\xi \in \R^{|\J|}$, $\xi \ge 0$ with at least one positive entry, such that
$$R_{\I\J} \s_\J = v^* {\bm 1} - \xi.$$
Recall $(\p,\q)$ is an NE.  By the assumption that $(\r,\s)$ is an NE, and the property of two-player zero-sum games, $(\p, \s)$ is also an NE with the same value $v^*$.
But $\p^\top R \s = \p_\I^\top R_{\I\J} \s_\J = v^* - \p_\I^\top \xi < v^*$, because all terms in $\p_\I$ are positive and at least one term in $\xi$ is positive.
This is a contradiction.
So case 2 will not happen.

Taken together, $\s=\q$.  Similarly, one can show $\r=\p$.

\textbf{Unique NE $\Rightarrow$ conditions:}
Let $(\p,\q)$ be the unique NE of $R$ with value $v^*$, and let $\I,\J$ be their support.

We first show SIISOW.
Equations~\eqref{eq:inpsame} and~\eqref{eq:inqsame} are immediate from NE definition.
Since $(\p,\q)$ is the only NE of the game, it satisfies Goldman and Tucker Corollary 3A.
The corollary states that 
\begin{eqnarray}
\forall i\in \mAa, (\e_i^\top R \q=v^*) \Rightarrow (i\in\I) \\
\forall j\in \mAb, (\p^\top R \e_j=v^*) \Rightarrow (j\in\J).
\end{eqnarray}
Their contraposition is
\begin{eqnarray}
\forall i\in \mAa, (i\notin\I) \Rightarrow (\e_i^\top R \q \neq v^*) \\
\forall j\in \mAb, (j\notin\J) \Rightarrow (\p^\top R \e_j \neq v^*).
\end{eqnarray}
But since $v^*$ is the NE game value, these imply
\begin{eqnarray}
\forall i\in \mAa, (i\notin\I) \Rightarrow (\e_i^\top R \q < v^*) \\
\forall j\in \mAb, (j\notin\J) \Rightarrow (\p^\top R \e_j < v^*).
\end{eqnarray}
Therefore, $(\p,\q)$ satisfies the SIISOW condition.

We next show invertability by contradition.
Suppose the matrix in Definition~\ref{df:inv} is not invertable.
Then either (i) $|\I|<|\J|$, (ii) $|\I|>|\J|$, or (iii) $|\I|=|\J|\ge 2$.
Case (iii) is due to the fact that should $|\I|=|\J|=1$, $R_{\I\J}$ is a scalar and the matrix 
$\begin{bmatrix} R_{\I\J} & -1 \\ 1 & 0 \end{bmatrix}$ with determinant 1 is always invertible.
We show that any one of the three cases leads to a second NE, contradicting the uniqueness of $(\p,\q)$.
In what follows we give the proof for (i) or (iii); case (ii) is similar to (i) but with respect to $R_{\I\J}^\top$ and $\p$, and is omitted.

In cases (i) or (iii) the following homogeneous linear system has a nonzero solution:
\begin{equation}
\label{eq:hls}
\begin{bmatrix} R_{\I\J} & -{\bm 1}_{|\I|} \\ {\bm 1}_{|\J|}^\top & 0 \end{bmatrix} 
\begin{bmatrix} \delta \\ x \end{bmatrix} 
 = 0,
\end{equation}
where $\delta \in \R^{|\J|}, x \in \R$.
This nonzero solution $(\delta,x)$ has some useful properties:
\begin{itemize}

\item 
$\delta$ sums to zero:
\begin{equation}
{\bm 1}^\top \delta =0.
\end{equation}
This follows directly from the second equality of~\eqref{eq:hls}.

\item 
$\delta\neq 0.$
This follows from the first equality of~\eqref{eq:hls}
\begin{equation}
\label{eq:RIJdelta}
R_{\I\J} \delta = x {\bm 1},
\end{equation}
otherwise both $\delta$ and $x$ would be zero, contradicting a nonzero solution.

\item 
$x=0$ and
\begin{equation}
\label{eq:RIJdelta0}
R_{\I\J} \delta = \bm 0.
\end{equation}
We first show $x=0$. Consider
\begin{eqnarray}
&& \p^\top R \begin{bmatrix} \delta \\ {\bm 0}_{|\mAb|-|\J|} \end{bmatrix} \label{eq:pRdelta0}\\
&=& \sum_{j\in\J} \p^\top R \e_j \delta_j \\ 
&=& \sum_{j\in\J} v^* \delta_j \\ 
&=& 0,
\end{eqnarray}
where the second equality follows from the SIISOW condition $\p^\top R \e_j = \p^\top R \q = v^*, \; \forall j\in\J$.
But at the same time, by the support of $\p$
\begin{eqnarray}
&& \p^\top R \begin{bmatrix} \delta \\ {\bm 0}_{|\mAb|-|\J|} \end{bmatrix} \\
&=& \p_{\I}^\top R_{\I\J} \delta \\
&=& \p_{\I}^\top x \bm 1 = x.
\end{eqnarray}
Therefore $x=0$.  Then use~\eqref{eq:RIJdelta} to obtain~\eqref{eq:RIJdelta0}.
\end{itemize}
We use this $\delta$ to construct another NE with the following steps:
\begin{enumerate}
\item We scale $\delta$ so its magnitude is sufficiently small.  The desired scale is determined by two constants:
  \begin{enumerate}
  \item Since we are under cases (i) or (iii), $|\J|\ge 2$. Thus the entries of $\q_\J$ cannot be 0 or 1: $\exists c_1>0: c_1\le q_j \le 1-c_1, \forall j\in\J$.
  \item By the SIISOW condition, 
$\e_i^\top R \q< v^*$ for $i\notin\I$.
Let $c_2 = v^*-\max_{i\notin\I} \e_i^\top R \q$.
  \end{enumerate}
We choose the scale 
\begin{equation}
c=\min\left({\frac{c_1}{\|\delta\|_\infty}}, \min_{i\notin I} {\frac{c_2}{|R_{i\J} \delta|}}\right).
\end{equation}
\item Set $\r = \q + \begin{bmatrix} c\delta \\ 0 \end{bmatrix}$.
\end{enumerate}
We claim $(\p,\r)$ is another NE:
\begin{itemize}
\item
Since $\delta$ sums to zero, $\q_\J+c\delta$ remains normalized; since $c\le {\frac{c_1}{\|\delta\|_\infty}}$ , all entries of $\q_\J+c\delta$ remains in $[0,1]$.
Therefore $\r \in \Delta_{\mAb}$ is a proper strategy.

\item $\r$ is a best response to $\p$:
\begin{eqnarray}
\p^\top R \r = \p^\top R \q +  \p^\top R \begin{bmatrix} c\delta \\ 0 \end{bmatrix} = v^*,
\end{eqnarray}
where we used~\eqref{eq:pRdelta0}.
Therefore, $\p^\top R \r = v^* \le \p^\top R \q', \forall \q' \in \Delta_{\mAb}$ because $\p$ is part of an NE.

\item
$\p$ is a best response to $\r$:
$\forall \p' \in \Delta_{\mAa}$, 
\begin{eqnarray}
&& \p'^\top R \r \\
&=& \sum_{i\in\I} p'_i \e_i^\top R \q
   + {\p'_\I}^\top R_{\I\J} c\delta
   +\sum_{i\notin\I} p'_i \left(\e_i^\top R \q + \e_i^\top R \begin{bmatrix} c\delta \\ 0 \end{bmatrix}\right) \nonumber \\
&=& \sum_{i\in\I} p'_i \e_i^\top R \q
   +\sum_{i\notin\I} p'_i \left(\e_i^\top R \q + \e_i^\top R \begin{bmatrix} c\delta \\ 0 \end{bmatrix}\right) \nonumber \\
&=& \sum_{i\in\I} p'_i v^*
   +\sum_{i\notin\I} p'_i \left(\e_i^\top R \q + \e_i^\top R \begin{bmatrix} c\delta \\ 0 \end{bmatrix}\right) \nonumber \\
&\le& \sum_{i\in\I} p'_i v^*
   +\sum_{i\notin\I} p'_i \left(v^* - c_2 + \e_i^\top R \begin{bmatrix} c\delta \\ 0 \end{bmatrix}\right) \nonumber \\
&=& \sum_{i\in\I} p'_i v^*
   +\sum_{i\notin\I} p'_i \left(v^* - c_2 + c R_{i\J}\delta \right). 
\end{eqnarray}
where the second equality follows from~\eqref{eq:RIJdelta0}, the next two lines from SIISOW.
Because $c \le \min_{i\notin I} {\frac{c_2}{|R_{i\J} \delta|}}$,
\begin{eqnarray}
&& \p'^\top R \r \\
&\le& \sum_{i\in\I} p'_i v^* +\sum_{i\notin\I} p'_i \left(v^* - c_2 + c_2\right) = v^* = \p^\top R \r. \nonumber
\end{eqnarray}
\end{itemize}
Because $\delta \neq 0$, $\r \neq \q$.
Thus $(\p,\r)\neq (\p,\q)$ is indeed a second NE, contradicting uniqueness.

\end{proof}

\subsection{Proof of Lemma~\ref{lem:erps}} \label{app:erps}

\begin{proof}  \label{proof:erpspf} To show uniqueness, we check that the conditions in Lemma~\ref{lem:unique} is satisfied,
\begin{equation} \begin{aligned}
\e_{i}^\top R \q &= 0 = \p^\top R \q, \forall\; i \in \I,
\\ \e_{i}^\top R \q &= -1 < 0 = \p^\top R \q, \forall\; i \notin \I,
\\ \p^\top R \e_{j} &= 0 = \p^\top R \q, \forall\; j \in \J,
\\ \p^\top R \e_{j} &= 1 > 0 = \p^\top R \q, \forall\; j \notin \J,
\end{aligned} \end{equation}
and we have $\begin{bmatrix} R_{\I \J} & -{\bm 1}_{|\I|} \\ {\bm 1}^\top_{|J|} & 0 \end{bmatrix}$ is invertible. 

To simplify the notations, we omit the modulo $k$ operation for the indices of $\p$ and $\q$. Observe that
\begin{equation} \begin{aligned}
\e^\top_{i} R \q &= - \dfrac{c}{\p_{i} \q_{i+1}} \q_{i+1} + \dfrac{c}{\p_{i} \q_{i+2}} \q_{i+2} = 0, \forall\; i \in \I,
\\ \e^\top_{i} R \q &= \displaystyle\sum_{j \in \J} -1 \q_{j} = -1, \forall\; i \notin \I,
\end{aligned} \end{equation}
and similarly,
\begin{equation} \begin{aligned}
\p^\top R \e_{j} &= \dfrac{c}{\p_{j-2} \q_{j}} \p_{j-2} - \dfrac{c}{\p_{j-1} q_{j}} \p_{j-1} = 0, \forall\; j \in \J,
\\ \p^\top R \e_{j} &= \displaystyle\sum_{i \in \I} 1 \p_{i} = 1, \forall\; j \notin \J.
\end{aligned} \end{equation}
In addition, we have,
\begin{equation} \begin{aligned}
\p^\top R \q &= \displaystyle\sum_{i \in \I} \p_{i} \left(\e^\top_{i} R \q\right) = 0.
\end{aligned} \end{equation}
Therefore, the SIISOW conditions are satisfied.

We now turn to the invertibility condition. For $k = 1, \begin{bmatrix} 0 & -1 \\ 1 & 0 \end{bmatrix}$ is invertible. For fixed $\p, \q$, for $k = 2$, we have,
\begin{equation} \begin{aligned}
& \det \begin{bmatrix} \dfrac{c}{\p_{0} \q_{0}} & - \dfrac{c}{\p_{0} \q_{1}} & -1 \\ - \dfrac{c}{\p_{1} \q_{0}} & \dfrac{c}{\p_{1} \q_{1}} & -1 \\ 1 & 1 & 0 \end{bmatrix}
\\ &= \det \begin{bmatrix} \dfrac{1}{\p_{0}} & 0 & 0 \\ 0 & \dfrac{1}{\p_{1}} & 0 \\ 0 & 0 & \dfrac{1}{c} \end{bmatrix} \det \begin{bmatrix} 1 & -1 & \p_{0} \\ -1 & 1 & \p_{1} \\ \q_{0} & \q_{1} & 0 \end{bmatrix} \det \begin{bmatrix} \dfrac{1}{\q_{0}} & 0 & 0 \\ 0 & \dfrac{1}{\q_{1}} & 0 \\ 0 & 0 & \dfrac{1}{c} \end{bmatrix}
\\ &= c \left(\p_{0} + \p_{1}\right) \dfrac{\q_{0} + \q_{1}}{\p_{0} \p_{1} \q_{0} \q_{1}}
\\ &> 0,
\end{aligned} \end{equation}
therefore it is invertible, similarly for $k = 3$,
\begin{equation} \begin{aligned}
& \det \begin{bmatrix} 0 & - \dfrac{c}{\p_{0} \q_{1}} & \dfrac{c}{\p_{0} \q_{2}} & -1 \\ \dfrac{c}{\p_{1} \q_{0}} & 0 & - \dfrac{c}{\p_{1} \q_{2}} & -1 \\ - \dfrac{c}{\p_{2} \q_{0}} & \dfrac{c}{\p_{2} \q_{1}} & 0 & -1 \\ 1 & 1 & 1 & 0 \end{bmatrix}
\\ &= \det \begin{bmatrix} \dfrac{1}{\p_{0}} & 0 & 0 & 0 \\ 0 & \dfrac{1}{\p_{1}} & 0 & 0 \\ 0 & 0 & \dfrac{1}{\p_{2}} & 0 \\ 0 & 0 & 0 & \dfrac{1}{c} \end{bmatrix} \det \begin{bmatrix} 0 & -1 & 1 & -\p_{0} \\ 1 & 0 & -1 & -\p_{1} \\ -1 & 1 & 0 & -\p_{2} \\ \q_{0} & \q_{1} & \q_{2} & 0 \end{bmatrix}  \det \begin{bmatrix} \dfrac{1}{\q_{0}} & 0 & 0 & 0 \\ 0 & \dfrac{1}{\q_{1}} & 0 & 0 \\ 0 & 0 & \dfrac{1}{\q_{2}} & 0 \\ 0 & 0 & 0 & \dfrac{1}{c} \end{bmatrix}
\\ &= c^{2} \dfrac{\left(\p_{0} + \p_{1} + \p_{2}\right) \left(\q_{0} + \q_{1} + \q_{2}\right)}{\p_{0} \p_{1} \p_{2} \q_{0} \q_{1} \q_{2}}
\\ &> 0,
\end{aligned} \end{equation}
and for $k = 4$,
\begin{equation} \begin{aligned}
& \det \begin{bmatrix} 0 & - \dfrac{c}{\p_{0} \q_{1}} & \dfrac{c}{\p_{0} \q_{2}} & 0 & -1 \\ 0 & 0 & - \dfrac{c}{\p_{1} \q_{2}} & \dfrac{c}{\p_{1} \q_{3}} & -1 \\ \dfrac{c}{\p_{2} \q_{0}} & 0 & 0 & - \dfrac{c}{\p_{2} \q_{3}} & -1 \\ - \dfrac{c}{\p_{3} \q_{0}} & \dfrac{c}{\p_{3} \q_{1}} & 0 & 0 & -1 \\ 1 & 1 & 1 & 1 & 0 \end{bmatrix}
\\ &= \det \begin{bmatrix} \dfrac{1}{\p_{0}} & 0 & 0 & 0 & 0 \\ 0 & \dfrac{1}{\p_{1}} & 0 & 0 & 0 \\ 0 & 0 & \dfrac{1}{\p_{2}} & 0 & 0 \\ 0 & 0 & 0 & \dfrac{1}{\p_{3}} & 0 \\ 0 & 0 & 0 & 0 & \dfrac{1}{c} \end{bmatrix} \det \begin{bmatrix} 0 & -1 & 1 & 0 & -\p_{0} \\ 0 & 0 & -1 & 1 & -\p_{1} \\ 1 & 0 & 0 & -1 & -\p_{2} \\ -1 & 1 & 0 & 0 & -\p_{3} \\ \q_{0} & \q_{1} & \q_{2} & \q_{3} & 0 \end{bmatrix} \det \begin{bmatrix} \dfrac{1}{\q_{0}} & 0 & 0 & 0 & 0 \\ 0 & \dfrac{1}{\q_{1}} & 0 & 0 & 0 \\ 0 & 0 & \dfrac{1}{\q_{2}} & 0 & 0 \\ 0 & 0 & 0 & \dfrac{1}{\q_{3}} & 0 \\ 0 & 0 & 0 & 0 & \dfrac{1}{c} \end{bmatrix}
\\ &= c^{3} \dfrac{\left(\p_{0} + \p_{1} + \p_{2} + \p_{3}\right) \left(\q_{0} + \q_{1} + \q_{2} + \q_{3}\right)}{\p_{0} \p_{1} \p_{2} \p_{3} \q_{0} \q_{1} \q_{2} \q_{3}}
\\ &> 0,
\end{aligned} \end{equation}
and in general, we can write $\begin{bmatrix} R_{\I \J} & -{\bm 1}_{\left| \I \right|} \\ {\bm 1}^\top_{\left| \J \right|} & 0 \end{bmatrix}$ as the product of $\text{\;diag\;} \left(\dfrac{1}{\p_{1}}, \dfrac{1}{\p_{2}}, ..., \dfrac{1}{\p_{k}}, \dfrac{1}{c}\right), \begin{bmatrix} R' & \p \\ \q^\top & 0 \end{bmatrix}$ , and $\text{\;diag\;} \left(\dfrac{1}{\q_{1}}, \dfrac{1}{\q_{2}}, ..., \dfrac{1}{\q_{k}}, \dfrac{1}{c}\right)$, where $R'$ is a matrix with entries,
\begin{equation} \begin{aligned}
R'_{ij} &= \begin{cases} -1 & \text{\;if\;} j = \left(i + 1\right) \mod k \\ 1 & \text{\;if\;} j = \left(i + 2\right) \mod k \\ 0 & \text{\;otherwise\;} \\ \end{cases} ,
\end{aligned} \end{equation}
with the above examples provided for $k = 2, 3, 4$,
\\* and the determinant is given by,
\begin{equation} \begin{aligned}
& \det \begin{bmatrix} R_{\I \J} & -{\bm 1}_{\left| \I \right|} \\ {\bm 1}^\top_{\left| \J \right|} & 0 \end{bmatrix}
\\ &= \det \text{\;diag\;} \left(\dfrac{1}{\p_{1}}, \dfrac{1}{\p_{2}}, ..., \dfrac{1}{\p_{k}}, \dfrac{1}{c}\right) \det \begin{bmatrix} R' & \p \\ \q^\top & 0 \end{bmatrix} \det \text{\;diag\;} \left(\dfrac{1}{\q_{1}}, \dfrac{1}{\q_{2}}, ..., \dfrac{1}{\q_{k}}, \dfrac{1}{c}\right)
\\ &= c^{k - 1} \dfrac{\displaystyle\sum_{i=1}^{k} \p_{i} \displaystyle\sum_{j=1}^{k} \q_{j}}{\displaystyle\prod_{i=1}^{k} \p_{i} \displaystyle\prod_{j=1}^{k} \q_{j}}
\\ &> 0.
\end{aligned} \end{equation}
This verifies the INV condition and completes the proof.
\end{proof}

\subsection{The Markov Game Modification Problem as An Optimization Problem} \label{app: lp}

Here we instantiate the general Game Modification problem (Definition~\ref{def:GM}) to Markov games as an optimization problem. 

\begin{df} [Game Modification for Two-Player Zero-Sum Markov Game] \label{df:gmmg} 
Given the cost function $\ell$, the target policy $\left(\p, \q\right)$ with supports $\I, \J$, target value range $\left[\underline{v}, \overline{v}\right]$, the game modification for Markov games can be written as the following optimization problem,
\begin{equation} \begin{aligned}
\label{eq:MGMO}
\displaystyle\inf_{R,v,\mQ} & \;\ell\left(R, R^{\circ}\right)
\\ \st & \left[\mQ_{h}\left(s\right)\right]_{\I_{h}\left(s\right)\bullet} \q_{h}\left(s\right) = v_{h}\left(s\right) {\bm 1}_{\left|\I_{h}\left(s\right)\right|}
\\ & \hspace{2em} \forall\; h \in \left[H\right], s \in \mathcal{S}
\\ & \p^\top_{h}\left(s\right) \left[\mQ_{h}\left(s\right)\right]_{\bullet\J_{h}\left(s\right)} = v_{h}\left(s\right) {\bm 1}^\top_{\left|\J_{h}\left(s\right)\right|}
\\ & \hspace{2em} \forall\; h \in \left[H\right], s \in \mathcal{S}
\\ & \left[\mQ_{h}\left(s\right)\right]_{\mAa \setminus \I_{h}\left(s\right)\bullet} \q_{h}\left(s\right) < v_{h}\left(s\right) {\bm 1}_{\left|\mAa \setminus \I_{h}\left(s\right)\right|}
\\ & \hspace{2em} \forall\; h \in \left[H\right], s \in \mathcal{S}
\\ & \p^\top_{h}\left(s\right) \left[\mQ_{h}\left(s\right)\right]_{\bullet\mAb \setminus \J_{h}\left(s\right)}  > v_{h}\left(s\right) {\bm 1}^\top_{\left|\mAb \setminus \J_{h}\left(s\right)\right|}
\\ & \hspace{2em} \forall\; h \in \left[H\right], s \in \mathcal{S}
\\ & \sigma_{\displaystyle\min}\left(\begin{bmatrix} \left[\mQ_{h}\left(s\right)\right]_{\I_{h}\left(s\right) \J_{h}\left(s\right)} & -{\bm 1}_{\left| \I_{h}\left(s\right) \right|} \\ {\bm 1}^\top_{\left| \J_{h}\left(s\right) \right|} & 0 \end{bmatrix}\right) > 0
\\ & \hspace{2em} \forall\; h \in \left[H\right], s \in \mathcal{S}
\\ & \mQ_{h}\left(s\right) = R_{h}\left(s\right) + \displaystyle\sum_{s' \in \mathcal{S}} P_{h}\left(s' | s\right) v_{h+1}\left(s'\right)
\\ & \hspace{2em} \forall\; h \in \left[H - 1\right], s \in \mathcal{S}
\\ & \mQ_{H}\left(s\right) = R_{H}\left(s\right), \forall\; s \in \mathcal{S}
\\ & \underline{v} \leq \displaystyle\sum_{s \in \mathcal{S}} P_{0}\left(s\right) v_{1}\left(s\right) \leq \overline{v}
\\ & -b \leq \left[R_{h}\left(s\right)\right]_{i j} \leq b
\\ & \hspace{2em} \forall\; \left(i, j\right) \in \mA, h \in \left[H\right], s \in \mathcal{S} .
\end{aligned} \end{equation}\end{df}

\subsection{Proof of Theorem~\ref{thm:feasible} and Theorem~\ref{cor:gamefeas}} \label{app: feas}

Theorem~\ref{thm:feasible} concerns the feasibility of modifying normal-form games in Proposition~\ref{thm:gmnf}, and Theorem~\ref{cor:gamefeas} concerns the feasibility of modifying $H$-period Markov games in Definition~\ref{df:gmmg}. Below we prove Theorem~\ref{cor:gamefeas}, from which Theorem~\ref{thm:feasible} follows as an special case with $H = 1$.

\textbf{Direction $\Rightarrow$.}
If $\pi = \left(\p, \q\right)$ is the unique Nash in stage game in period $h \in \left[H\right]$, state $s \in \mathcal{S}$, then by Lemma~\ref{lem:unique}, $\begin{bmatrix} R_{\I_{h}\left(s\right) \J_{h}\left(s\right)} & -{\bm 1}_{\I_{h}\left(s\right)} \\ {\bm 1}^\top_{\J_{h}\left(s\right)} & 0 \end{bmatrix}$ is an invertible square matrix, therefore, $\left| \I_{h}\left(s\right) \right| = \left| \J_{h}\left(s\right) \right|$.
\\* Now to show that $\left(-H b, H b\right) \cap \left[\underline{v}, \overline{v}\right] =$ empty leads to infeasibility, note that either,
\begin{equation} \begin{aligned}
\overline{v} &\geq H b,
\end{aligned} \end{equation}
or,
\begin{equation} \begin{aligned}
\underline{v} &\leq - H b,
\end{aligned} \end{equation}
meaning the value of at least one stage game at least $b$ or at most $-b$, and the SIISOW conditions imply that there are some entries of $R_{h}\left(s\right)$ that are strictly larger than $b$ or strictly smaller than $-b$, which contradicts the reward bound conditions.

\textbf{Direction $\Leftarrow$.}
Fix a stage game in period $h \in \left[H\right]$, state $s \in \mathcal{S}$, if $\left| \I_{h}\left(s\right) \right| = \left| \J_{h}\left(s\right) \right| = k$ for some $k$, then without loss of generality, we can rename the actions so that $\I_{h}\left(s\right) = \J_{h}\left(s\right) = \left\{0, 1, 2, ..., k - 1\right\}$ and Lemma~\ref{lem:erps} provides a game with the unique Nash equilibrium $\left(\p_{h}\left(s\right), \q_{h}\left(s\right)\right)$. Note that since the value of $R^{\textup{\;eRPS\;}}$ is $0$, all stage games have value $0$, so we have, for every $h \in \left[H\right], s \in \mathcal{S}$,
\begin{equation} \begin{aligned}
\mQ_{h}\left(s\right) &= R_{h}\left(s\right).
\end{aligned} \end{equation}
The $\left(-H b, H b\right) \cap \left[\underline{v}, \overline{v}\right] \neq \emptyset$  condition guarantees the existence of some $v^\star \in \left[\underline{v}, \overline{v}\right]$ that satisfies,
\begin{equation} \begin{aligned}
-H b &< v^\star < H b.
\end{aligned} \end{equation}
Now consider the Markov game $\left(R, P\right)$ with rewards defined by,
\begin{equation} \begin{aligned}
R_{h}\left(s\right) &= \delta R^{\textup{\;eRPS\;} \left(\p_{h}\left(s\right), \q_{h}\left(s\right)\right)} + \dfrac{1}{H} v^\star.
\end{aligned} \end{equation}
This implies that the $Q$ matrices can be computed as recursively for $h = H-1, H-2, ..., 1$,
\begin{equation} \begin{aligned}
v_{h}\left(s\right) &= \dfrac{H-h+1}{H} v^\star,
\\ \mQ_{h}\left(s\right) &= R_{h}\left(s\right) + \displaystyle\sum_{s' \in \mathcal{S}} P_{h}\left(s' | s\right) v_{h+1}\left(s\right)
\\ &= R_{h}\left(s\right) + \displaystyle\sum_{s' \in \mathcal{S}} P_{h}\left(s' | s\right) \dfrac{H-h}{H} v^\star
\\ &= R_{h}\left(s\right) + \dfrac{H-h}{H} v^\star
\\ &= \delta R^{\textup{\;eRPS\;} \left(\p_{h}\left(s\right), \q_{h}\left(s\right)\right)} + \dfrac{H-h+1}{H} v^\star,
\end{aligned} \end{equation}
which is an affine transformation of $R^{\textup{\;eRPS\;}}$, so it has unique Nash $\left(\p_{h}\left(s\right), \q_{h}\left(s\right)\right)$ with value $\dfrac{H-h+1}{H} v^\star$. In particular, the value of this game is given by,
\begin{equation} \begin{aligned}
v_{0} &\coloneqq \displaystyle\sum_{s \in \mathcal{S}} P_{0}\left(s\right) v_{1}\left(s\right)
\\ &= \displaystyle\sum_{s \in \mathcal{S}} P_{0}\left(s\right) \dfrac{H-1+1}{H} v^\star
\\ &= v^\star,
\end{aligned} \end{equation}
which satisfies the value range constraint. In addition, for $\delta$ sufficiently small, the entry bound conditions are satisfied as well. In particular, if $\delta < \min\left\{H b - v^\star, H b + v^\star\right\}$, for which the righthand side is strictly positive due to the condition $\left(-H b, H b\right) \cap \left[\underline{v}, \overline{v}\right] \neq \emptyset$, we have $R_{h}\left(s\right) \in \left[-b, b\right]$.

\subsection{Proof of Feasibility/Optimality for RAP and RAP-MG Algorithms (Theorem~\ref{thm:perturb} and Theorem~\ref{cor:perturbmg})} \label{app:opt}

Theorem~\ref{thm:perturb} concerns the feasibility and optimality of the RAP algorithm for normal form games. This result is a special case of Theorem~\ref{cor:perturbmg} below for the RAP-MG algorithm for Markov games.

\begin{proof}  \label{proof:perturbpf}

We show the general result for $H$-period Markov games, and Theorem~\ref{thm:perturb} is the special case when $H = 1$.

\textbf{Existence.} Existence of a solution is implied by Theorem~\ref{cor:gamefeas} with value bounds $\left[- H b + H \lambda, H b - H \lambda\right]$, and due to~\eqref{eq:li}, we have,
\begin{equation} \begin{aligned}
\left(- H b + H \lambda, H b - H \lambda\right) \cap \left[\underline{v}, \overline{v}\right] &\neq \emptyset,
\end{aligned} \end{equation}
and therefore, Theorem~\ref{cor:gamefeas} implies the feasible of the problem thus existence of a solution.

\textbf{Feasibility.} We only have to check the INV constraints since $\iota, \lambda > 0$ implies that the other constraints in the original problem are satisfied. We check that for every stage game $\mQ$ in period $h \in \left[H\right], s \in \mathcal{S}$, we have $\mQ_{h}\left(s\right) = \mQ'_{h}\left(s\right) + \varepsilon R^{\textup{\;eRPS\;} \left(\p_{h}\left(s\right), \q_{h}\left(s\right)\right)}$ satisfies INV, where $\mQ'_{h}\left(s\right)$ is the solution to the optimization. To simplify the notations, we drop the $\left(h, s\right)$ indices.

We use the following properties of $R^{\textup{\;eRPS\;}}$ from the proof of Lemma~\ref{lem:erps},
\begin{equation} \begin{aligned}
R^{\textup{\;eRPS\;}}_{\I\bullet} \q &= {\bm 0}_{\left| \I \right|},
\\ \p^\top R^{\textup{\;eRPS\;}}_{\bullet\J} &= {\bm 0}_{\left| \J \right|},
\\ R^{\textup{\;eRPS\;}}_{\mAa\setminus\I\bullet} \q &= -{\bm 1}_{\left| \mAa\setminus\I \right|},
\\ \p^\top R^{\textup{\;eRPS\;}}_{\bullet\mAb\setminus\J} &= {\bm 1}_{\left| \mAb\setminus\J \right|}.
\end{aligned} \end{equation}
Now we check the three conditions of the attacker's problem are satisfied. We have
\begin{equation} \begin{aligned}
\mQ_{\I\bullet} \q &= \mQ'_{\I} \q + \varepsilon R^{\textup{\;eRPS\;}}_{\I\bullet} \q
\\ &= v {\bm 1}_{\left| \I \right|}
\\ &= v' {\bm 1}_{\left| \I \right|},
\end{aligned} \end{equation}
and similarly,
\begin{equation} \begin{aligned}
\p^\top \mQ_{\bullet\J} &= \p^\top \mQ'_{\bullet\J} + \varepsilon \p R^{\textup{\;eRPS\;}}_{\bullet\J}
\\ &= v {\bm 1}_{\left| \J \right|}
\\ &= v' {\bm 1}_{\left| \J \right|}.
\end{aligned} \end{equation}
We also have 
\begin{equation} \begin{aligned}
\mQ_{\mAa\setminus\I\bullet} \q &= \mQ'_{\mAa\setminus\I\bullet} \q + \varepsilon R^{\textup{\;eRPS\;}}_{\mAa\setminus\I} \q
\\ &< v {\bm 1}_{\left| \mAa\setminus\I \right|} - \varepsilon {\bm 1}_{\left| \mAa\setminus\I \right|}
\\ &< v' {\bm 1}_{\left| \mAa\setminus\I \right|}.
\end{aligned} \end{equation}
and similarly,
\begin{equation} \begin{aligned}
\p^\top \mQ_{\bullet\mAb\setminus\J} &= \p^\top \mQ'_{\bullet\mAb\setminus\J} + \varepsilon R^{\textup{\;eRPS\;}}_{\mAb\setminus\J} \q
\\ &> v {\bm 1}_{\left| \mAb\setminus\J \right|} + \varepsilon {\bm 1}_{\left| \mAb\setminus\J \right|}
\\ &> v' {\bm 1}_{\left| \mAb\setminus\J \right|}.
\end{aligned} \end{equation}
Next we show that $\begin{bmatrix} \mQ_{\I\J} & -{\bm 1}_{\left| \I \right|} \\ {\bm 1}^\top_{\left| \J \right|} & 0 \end{bmatrix}$ is invertible with probability $1$, in particular, since $\begin{bmatrix} R^{\textup{\;eRPS\;}} & -{\bm 1}_{\left| \I \right|} \\ {\bm 1}^\top_{\left| \J \right|} & 0 \end{bmatrix}$ is invertible by Lemma~\ref{lem:erps}, we can write its singular value decomposition,
\begin{equation} \begin{aligned}
\begin{bmatrix} R^{\textup{\;eRPS\;}} & -{\bm 1}_{\left| \I \right|} \\ {\bm 1}^\top_{\left| \J \right|} & 0 \end{bmatrix} &= U \Sigma V^\top ,
\end{aligned} \end{equation}
for some orthonormal $U, V \in \mathbb{R}^{\left(\left| \I \right| + 1\right) \times \left(\left| \J \right| + 1\right)}$ and nonsingular diagonal matrix $\Sigma \in \mathbb{R}^{\left(\left| \I \right| + 1\right) \times \left(\left| \J \right| + 1\right)}$. Consider the event $\begin{bmatrix} \mQ_{\I \J} & -{\bm 1}_{\left| \I \right|} \\ {\bm 1}^\top_{\left| \J \right|} & 0 \end{bmatrix}$ is singular. Then $\begin{bmatrix} \mQ_{\I \J} & - \left(1 + \varepsilon\right) {\bm 1}_{\left| \I \right|} \\ \left(1 + \varepsilon\right) {\bm 1}^\top_{\left| \J \right|} & 0 \end{bmatrix}$ is singular, and the following matrix is also singular:
\begin{equation} \begin{aligned}
& \Sigma^{- 1/2} U^\top \begin{bmatrix} \mQ_{\I \J} & - \left(1 + \varepsilon\right) {\bm 1}_{\left| \I \right|} \\ \left(1 + \varepsilon\right) {\bm 1}^\top_{\left| \J \right|} & 0 \end{bmatrix} V \Sigma^{- 1/2}
\\ &= \Sigma^{- 1/2} U^\top \begin{bmatrix} \mQ'_{\I \J} + \varepsilon R^{\textup{\;eRPS\;}} & - \left(1 + \varepsilon\right) {\bm 1}_{\left| \I \right|} \\ \left(1 + \varepsilon\right) {\bm 1}^\top_{\left| \J \right|} & 0 \end{bmatrix} V \Sigma^{- 1/2}
\\ &= \Sigma^{- 1/2} U^\top \begin{bmatrix} \mQ'_{\I \J} + \dfrac{\varepsilon'}{1-\varepsilon'} R^{\textup{\;eRPS\;}} & - \dfrac{1}{1-\varepsilon'} {\bm 1}_{\left| \I \right|} \\ \dfrac{1}{1-\varepsilon'} {\bm 1}^\top_{\left| \J \right|} & 0 \end{bmatrix} V \Sigma^{- 1/2}
\\ & \hspace{2em} \text{\;where\;} \varepsilon' \coloneqq \dfrac{\varepsilon}{1+\varepsilon} = 1 - \dfrac{1}{1+\varepsilon},
\\ & \hspace{2em} \text{\;which implies\;} \varepsilon = \dfrac{1}{1-\varepsilon'}-1 = \dfrac{\varepsilon'}{1-\varepsilon'},
\\ &= \dfrac{1}{1-\varepsilon'} \Sigma^{- 1/2} U^\top \begin{bmatrix} \left(1-\varepsilon'\right) \mQ'_{\I \J} + \varepsilon' R^{\textup{\;eRPS\;}} & - {\bm 1}_{\left| \I \right|} \\ {\bm 1}^\top_{\left| \J \right|} & 0 \end{bmatrix} V \Sigma^{- 1/2}
\\ &= \Sigma^{- 1/2} U^\top \begin{bmatrix} \mQ'_{\I \J} & - {\bm 1}_{\left| \I \right|} \\ {\bm 1}^\top_{\left| \J \right|} & 0 \end{bmatrix} V \Sigma^{- 1/2}
\\ & \hspace{2em} + \dfrac{\varepsilon'}{1-\varepsilon'} \Sigma^{- 1/2} U^\top \begin{bmatrix} R^{\textup{\;eRPS\;}}_{\I \J} & - {\bm 1}_{\left| \I \right|} \\ {\bm 1}^\top_{\left| \J \right|} & 0 \end{bmatrix} V \Sigma^{- 1/2}
\\ &= \Sigma^{- 1/2} U^\top \begin{bmatrix} \mQ'_{\I \J} & - {\bm 1}_{\left| \I \right|} \\ {\bm 1}^\top_{\left| \J \right|} & 0 \end{bmatrix} V \Sigma^{- 1/2} + \varepsilon I.
\end{aligned} \end{equation}
Consequently,  there exists a nonzero vector $x \in \mathbb{R}^{\left| \I \right| + 1} = \mathbb{R}^{\left| \J \right| + 1}$ such that,
\begin{equation} \begin{aligned}
\Sigma^{- 1/2} U^\top \begin{bmatrix} \mQ'_{\I \J} & - {\bm 1}_{\left| \I \right|} \\ {\bm 1}^\top_{\left| \J \right|} & 0 \end{bmatrix} V \Sigma^{- 1/2} x &= - \varepsilon x .
\end{aligned} \end{equation}
This means that  $- \varepsilon$ is an eigenvalue of the following deterministic matrix,
\begin{equation} \begin{aligned}
&\Sigma^{- 1/2} U^\top \begin{bmatrix} \mQ'_{\I \J} & - {\bm 1}_{\left| \I \right|} \\ {\bm 1}^\top_{\left| \J \right|} & 0 \end{bmatrix} V \Sigma^{- 1/2},
\end{aligned} \end{equation}
which happens with probability $0$ since $\varepsilon \sim \text{\;Unif\;} \left[-\lambda, \lambda\right]$ is continuous.

\textbf{Optimality.} Fix $\varepsilon > 0$. Consider a feasible solution to~\eqref{eq:MGMO}, $\left(R^{\left(\varepsilon\right)}, v^{\left(\varepsilon\right)}\right)$, that satisfies
\begin{equation} \begin{aligned}
\ell\left(R^{\left(\varepsilon\right)}, R^{\circ}\right) - C^\star &< \dfrac{\varepsilon}{2} .
\end{aligned} \end{equation}
In particular, feasibility of $R^{\left(\varepsilon\right)}$ implies, for every $h \in \left[H\right], s \in \mathcal{S}$,
\begin{equation} \begin{aligned}
\left[\mQ^{\left(\varepsilon\right)}_{h}\left(s\right)\right]_{\I\bullet} \q &= v^{\left(\varepsilon\right)}_{h}\left(s\right) {\bm 1}_{\left| \I \right|}
\\ \p^\top \left[\mQ^{\left(\varepsilon\right)}_{h}\left(s\right)\right]_{\bullet\J} &= v^{\left(\varepsilon\right)}_{h}\left(s\right) {\bm 1}_{\left| \J \right|}^\top
\\ \left[\mQ^{\left(\varepsilon\right)}_{h}\left(s\right)\right]_{\mAa\setminus\I\bullet} \q &< v^{\left(\varepsilon\right)}_{h}\left(s\right) {\bm 1}_{\left| \mAa\setminus\I \right|}
\\ \p^\top \left[\mQ^{\left(\varepsilon\right)}_{h}\left(s\right)\right]_{\bullet\mAb\setminus\J} &> v^{\left(\varepsilon\right)}_{h}\left(s\right) {\bm 1}_{\left| \mAb\setminus\J \right|}
\\ \sigma_{\displaystyle\min}\left(\begin{bmatrix} \mQ^{\left(\varepsilon\right)}_{h}\left(s\right) & _{\I\J} \\ -{\bm 1}_{\left| \I \right|} & {\bm 1}_{\left| \J \right|} \end{bmatrix} 0\right) &> 0
\\ \mQ^{\left(\varepsilon\right)}_{h}\left(s\right) &= R^{\left(\varepsilon\right)}_{h}\left(s\right) + \displaystyle\sum_{s' \in \mathcal{S}} P_{h}\left(s'|s\right) v^{\left(\varepsilon\right)}_{h+1}\left(s'\right)
\\ -b &\leq \left[R^{\left(\varepsilon\right)}_{h}\left(s\right)\right]_{ij} \leq b, \forall\; \left(i,j\right) \in \mA.
\end{aligned} \end{equation}
Due to the strict SOW inequality in~\eqref{eq:MGMO}, we can find the $\iota^{\left(\varepsilon\right)} > 0$ such that the SOW conditions in~\eqref{eq:MGMLP} is also satisfied,
\begin{equation} \begin{aligned}
\iota^{\left(\varepsilon\right)} &\coloneqq \displaystyle\min_{h \in \left[H\right], s \in \mathcal{S}}\left\{v^{\left(\varepsilon\right)}_{h}\left(s\right) {\bm 1}_{\left| \mAa\setminus\I \right|} - \left[\mQ^{\left(\varepsilon\right)}_{h}\left(s\right)\right]_{\mAa\setminus\I\bullet} \q, \p^\top \left[\mQ^{\left(\varepsilon\right)}_{h}\left(s\right)\right]_{\bullet\mAb\setminus\J} - v^{\left(\varepsilon\right)}_{h}\left(s\right) {\bm 1}_{\left| \mAb\setminus\J \right|}\right\},
\end{aligned} \end{equation}
where the $\displaystyle\min$ is element-wise for the vectors.

Since $v^{\left(\varepsilon\right)} \in \left(-H b, H b\right)$, we can find the value gap $\lambda^{\left(\varepsilon\right)} > 0$,
\begin{equation} \begin{aligned}
\lambda^{\left(\varepsilon\right)}&\coloneqq b - \displaystyle\min_{h \in \left[H\right], s \in \mathcal{S}, \left(i,j\right) \in \mA} \left| v_{h}\left(s\right) - P_{ij}\left(s'|s\right) v_{h+1}\left(s'\right) \right|,
\end{aligned} \end{equation}
by noting that if $\lambda^{\left(\varepsilon\right)} = 0$, then $\left| v^{\left(\varepsilon\right)} \right| \geq H b$ which contradicts our assumption.

Now we define the following $\delta$,
\begin{equation} \begin{aligned}
\delta &\coloneqq \displaystyle\min\left\{\dfrac{\iota^{\left(\varepsilon\right)}}{2} , \dfrac{\varepsilon \lambda^{\left(\varepsilon\right)}}{2 L b H \left(H + 1\right) \left| \mathcal{S} \right| \left| \mA \right|}\right\}.
\end{aligned} \end{equation}
Note that $R^{\left(\varepsilon\right)}$ does not satisfy~\eqref{eq:MGMLP} due the tighter bounds on the entries, meaning $-b + \lambda \leq \left[R^{\left(\varepsilon\right)}_{h}\left(s\right)\right]_{ij} \leq b - \lambda$ may not be satisfied for some $h \in \left[H\right], s \in \mathcal{S}, \left(i,j\right) \in \mA$. We define $R'^{\left(\varepsilon\right)}$ as follows and show that $\left(R'^{\left(\varepsilon\right)}, v^{\left(\varepsilon\right)}\right)$ is feasible to~\eqref{eq:MGMLP}, for every $h \in \left[H\right], s \in \mathcal{S}, \left(i, j\right) \in \mA$,
\begin{equation} \begin{aligned}
\left[R'^{\left(\varepsilon\right)}_{h}\left(s\right)\right]_{ij} &\coloneqq \begin{cases} \left(1 - \dfrac{\delta}{\lambda^{\left(\varepsilon\right)}}\right) \left[\mQ^{\left(\varepsilon\right)}_{h}\left(s\right)\right]_{ij} + \dfrac{v^{\left(\varepsilon\right)}_{h}\left(s\right) \delta}{\lambda^{\left(\varepsilon\right)}} - \displaystyle\sum_{s' \in \mathcal{S}} \left[P_{h}\left(s'|s\right)\right]_{ij} v^{\left(\varepsilon\right)}_{h+1}\left(s'\right) & \text{\;if\;} i \in \I_{h}\left(s\right), j \in \J_{h}\left(s\right) \\ \displaystyle\min\left\{\displaystyle\max\left\{\left[R^{\left(\varepsilon\right)}_{h}\left(s\right)\right]_{ij}, -b+\delta\right\},b-\delta\right\} & \text{\;otherwise\;} \\ \end{cases}.
\end{aligned} \end{equation}
In particular, we have for $i \in \I_{h}\left(s\right), j \in \J_{h}\left(s\right)$,
\begin{equation} \begin{aligned}
\left[\mQ'^{\left(\varepsilon\right)}_{h}\left(s\right)\right]_{ij} &= \left(1 - \dfrac{\delta}{\lambda^{\left(\varepsilon\right)}}\right) \left[\mQ^{\left(\varepsilon\right)}_{h}\left(s\right)\right]_{ij} + \dfrac{v^{\left(\varepsilon\right)}_{h}\left(s\right) \delta}{\lambda^{\left(\varepsilon\right)}} .
\end{aligned} \end{equation}
Now, to check the feasibility of $\left(R'^{\left(\varepsilon\right)}, v^{\left(\varepsilon\right)}\right)$ to~\eqref{eq:MGMLP}, fix $h \in \left[H\right], s \in \mathcal{S}$. To simplify the notations, we drop the $\left(h, s\right)$ indices. Observe that
\begin{equation} \begin{aligned}
\mQ'^{\left(\varepsilon\right)}_{\I\bullet} \q &= \left(\left(1 - \dfrac{\delta}{\lambda^{\left(\varepsilon\right)}}\right) \mQ^{\left(\varepsilon\right)}_{\I\bullet} + \dfrac{v^{\left(\varepsilon\right)} \delta}{\lambda^{\left(\varepsilon\right)}}\right) \q, \text{\;since\;} \q_{\mAb\setminus\J} = {\bm 0}_{\left| \mAb\setminus\J \right|}
\\ &= \left(1 - \dfrac{\delta}{\lambda^{\left(\varepsilon\right)}}\right) \mQ^{\left(\varepsilon\right)}_{\I\bullet} \q + \dfrac{v^{\left(\varepsilon\right)} \delta}{\lambda^{\left(\varepsilon\right)}} {\bm 1}_{\I\J} \q
\\ &= \left(1 - \dfrac{\delta}{\lambda^{\left(\varepsilon\right)}}\right) v^{\left(\varepsilon\right)} {\bm 1}_{\left| \I \right|} + \dfrac{v^{\left(\varepsilon\right)} \delta}{\lambda^{\left(\varepsilon\right)}} {\bm 1}_{\I\J} \q, \text{\;since\;} \left(R^{\left(\varepsilon\right)}, v^{\left(\varepsilon\right)}\right) \text{\;is feasible\;}
\\ &= \left(1 - \dfrac{\delta}{\lambda^{\left(\varepsilon\right)}}\right) v^{\left(\varepsilon\right)} + \dfrac{v^{\left(\varepsilon\right)} \delta}{\lambda^{\left(\varepsilon\right)}}
\\ &= v^{\left(\varepsilon\right)},
\end{aligned} \end{equation}
and similarly,
\begin{equation} \begin{aligned}
\p^\top \mQ'^{\left(\varepsilon\right)}_{\bullet\J} &= \p^\top \left(\left(1 - \dfrac{\delta}{\lambda^{\left(\varepsilon\right)}}\right) \mQ^{\left(\varepsilon\right)}_{\bullet\J} + \dfrac{v^{\left(\varepsilon\right)} \delta}{\lambda^{\left(\varepsilon\right)}}\right)
\\ &= \left(1 - \dfrac{\delta}{\lambda^{\left(\varepsilon\right)}}\right) v^{\left(\varepsilon\right)} + \dfrac{v^{\left(\varepsilon\right)} \delta}{\lambda^{\left(\varepsilon\right)}}
\\ &= v^{\left(\varepsilon\right)}.
\end{aligned} \end{equation}
Consider any $\iota < \delta$, we have,
\begin{equation} \begin{aligned}
\mQ'^{\left(\varepsilon\right)}_{\mAa\setminus\I\bullet} \q &\leq \left(\mQ^{\left(\varepsilon\right)}_{\mAa\setminus\I\bullet} + \dfrac{\iota^{\left(\varepsilon\right)}}{2}\right) \q
\\ &\leq \mQ^{\left(\varepsilon\right)}_{\mAa\setminus\I\bullet} \q + \dfrac{\iota^{\left(\varepsilon\right)}}{2} {\bm 1}_{\left| \mAa\setminus\I \right|}
\\ &\leq \left(v^{\left(\varepsilon\right)} - \iota^{\left(\varepsilon\right)}\right) {\bm 1}_{\left| \mAa\setminus\I \right|} + \dfrac{\iota^{\left(\varepsilon\right)}}{2} {\bm 1}_{\left| \mAa\setminus\I \right|}
\\ &\leq \left(v^{\left(\varepsilon\right)} - \dfrac{\iota^{\left(\varepsilon\right)}}{2}\right) {\bm 1}_{\left| \mAa\setminus\I \right|}
\\ &\leq \left(v^{\left(\varepsilon\right)} - \delta\right) {\bm 1}_{\left| \mAa\setminus\I \right|}
\\ &\leq \left(v^{\left(\varepsilon\right)} - \iota\right) {\bm 1}_{\left| \mAa\setminus\I \right|},
\end{aligned} \end{equation}
and similarly,
\begin{equation} \begin{aligned}
\p^\top \mQ'^{\left(\varepsilon\right)}_{\bullet\mAb\setminus\J} &\geq \p^\top \left(\mQ^{\left(\varepsilon\right)}_{\bullet\mAb\setminus\J} - \dfrac{\iota^{\left(\varepsilon\right)}}{2}\right)
\\ &\geq \left(v^{\left(\varepsilon\right)} + \iota^{\left(\varepsilon\right)}\right) {\bm 1}_{\left| \mAb\setminus\J \right|} - \dfrac{\iota^{\left(\varepsilon\right)}}{2} {\bm 1}_{\left| \mAb\setminus\J \right|}
\\ &\geq \left(v^{\left(\varepsilon\right)} + \dfrac{\iota^{\left(\varepsilon\right)}}{2}\right) {\bm 1}_{\left| \mAb\setminus\J \right|}
\\ &\geq \left(v^{\left(\varepsilon\right)} + \iota\right) {\bm 1}_{\left| \mAb\setminus\J \right|}.
\end{aligned} \end{equation}
Now to show that $\begin{bmatrix} \mQ'^{\left(\varepsilon\right)}_{\I\J} & -{\bm 1}_{\left| \J \right|} \\ {\bm 1}_{\left| \I \right|}^\top & 0 \end{bmatrix}$ is invertible, since $\begin{bmatrix} \mQ^{\left(\varepsilon\right)}_{\I\J} & -{\bm 1}_{\left| \J \right|} \\ {\bm 1}_{\left| \I \right|}^\top & 0 \end{bmatrix}$ is invertible, there exists vector $\begin{bmatrix} x \\ t \end{bmatrix} \neq {\bm 0}_{\left| \J \right|+1}$, such that
\begin{equation} \begin{aligned}
& \begin{bmatrix} \mQ^{\left(\varepsilon\right)}_{\I\J} & -{\bm 1}_{\left| \J \right|} \\ {\bm 1}_{\left| \I \right|}^\top & 0 \end{bmatrix} \begin{bmatrix} x \\ t \end{bmatrix} = {\bm 0}_{\left| \J \right|+1}
\\ &\Rightarrow \begin{cases} \mQ^{\left(\varepsilon\right)}_{\I\J} x - t {\bm 1}_{\left| \J \right|} = {\bm 0}_{\left| \J \right|} \\ {\bm 1}_{\left| \I \right|}^\top x = 0 \\ \end{cases}
\\ &\Rightarrow \begin{cases} \left(1 - \dfrac{\delta}{\lambda^{\left(\varepsilon\right)}}\right) \mQ^{\left(\varepsilon\right)}_{\I\J} x - \left(1 - \dfrac{\delta}{\lambda^{\left(\varepsilon\right)}}\right) t {\bm 1}_{\left| \J \right|} = {\bm 0}_{\left| \J \right|} \\ {\bm 1}_{\left| \I \right|}^\top x = 0 \\ \end{cases}
\\ &\Rightarrow \begin{cases} \left(1 - \dfrac{\delta}{\lambda^{\left(\varepsilon\right)}}\right) \mQ^{\left(\varepsilon\right)}_{\I\J} x + \dfrac{v^{\left(\varepsilon\right)} \delta}{\lambda^{\left(\varepsilon\right)}} {\bm 1}_{\I\J} x - \left(1 - \dfrac{\delta}{\lambda^{\left(\varepsilon\right)}}\right) t {\bm 1}_{\left| \J \right|} = {\bm 0}_{\left| \J \right|} \\ {\bm 1}_{\left| \I \right|}^\top x = 0 \\ \end{cases} , \text{\;since\;} {\bm 1}_{\left| \I \right|}^\top x = 0
\\ &\Rightarrow \begin{cases} \left(\left(1 - \dfrac{\delta}{\lambda^{\left(\varepsilon\right)}}\right) \mQ^{\left(\varepsilon\right)}_{\I\J} + \dfrac{v^{\left(\varepsilon\right)} \delta}{\lambda^{\left(\varepsilon\right)}}\right) x - \left(1 - \dfrac{\delta}{\lambda^{\left(\varepsilon\right)}}\right) t {\bm 1}_{\left| \J \right|} = {\bm 0}_{\left| \J \right|} \\ {\bm 1}_{\left| \I \right|}^\top x = 0 \\ \end{cases}
\\ &\Rightarrow \begin{cases} \mQ'^{\left(\varepsilon\right)}_{\I\J} x - \left(1 - \dfrac{\delta}{\lambda^{\left(\varepsilon\right)}}\right) t {\bm 1}_{\left| \J \right|} = {\bm 0}_{\left| \J \right|} \\ {\bm 1}_{\left| \I \right|}^\top x = 0 \\ \end{cases}
\\ &\Rightarrow \begin{bmatrix} \mQ'^{\left(\varepsilon\right)}_{\I\J} & -{\bm 1}_{\left| \J \right|} \\ {\bm 1}_{\left| \I \right|}^\top & 0 \end{bmatrix} \begin{bmatrix} x \\ \left(1 - \dfrac{\delta}{\lambda^{\left(\varepsilon\right)}}\right) t \end{bmatrix} = {\bm 0}_{\left| \J \right|+1}.
\end{aligned} \end{equation}
Since $\begin{bmatrix} x \\ \left(1 - \dfrac{\delta}{\lambda^{\left(\varepsilon\right)}}\right) t \end{bmatrix} \neq {\bm 0}_{\left| \J \right|+1}$, we have $\begin{bmatrix} \mQ'^{\left(\varepsilon\right)}_{\I\J} & -{\bm 1}_{\left| \J \right|} \\ {\bm 1}_{\left| \I \right|}^\top & 0 \end{bmatrix}$ is invertible.

Since we did not change the value $v^{\left(\varepsilon\right)}$, the value range constraint is still satisfied,
\begin{equation} \begin{aligned}
\underline{v} &\leq v^{\left(\varepsilon\right)} \leq \overline{v}.
\end{aligned} \end{equation}
For the range condition, we use the short-hand notation,
\begin{equation} \begin{aligned}
\Delta_{ij} v^{\left(\varepsilon\right)}_{h}\left(s\right) &\coloneqq v^{\left(\varepsilon\right)}_{h}\left(s\right) - \displaystyle\sum_{s' \in \mathcal{S}} \left[P_{h}\left(s'|s\right)\right]_{ij} v^{\left(\varepsilon\right)}_{h+1}\left(s'\right).
\end{aligned} \end{equation}
Note that we have,
\begin{equation} \begin{aligned}
R'^{\left(\varepsilon\right)}_{h}\left(s\right) &= \mQ'^{\left(\varepsilon\right)}_{h}\left(s\right) - \displaystyle\sum_{s' \in \mathcal{S}} P_{h}\left(s'|s\right) v^{\left(\varepsilon\right)}_{h+1}\left(s'\right)
\\ &= \left(1 - \dfrac{\delta}{\lambda^{\left(\varepsilon\right)}}\right) \mQ^{\left(\varepsilon\right)}_{h}\left(s\right) + v^{\left(\varepsilon\right)}_{h}\left(s\right) \dfrac{\delta}{\lambda^{\left(\varepsilon\right)}} - \displaystyle\sum_{s' \in \mathcal{S}} P_{h}\left(s'|s\right) v^{\left(\varepsilon\right)}_{h+1}\left(s'\right)
\\ &= \left(1 - \dfrac{\delta}{\lambda^{\left(\varepsilon\right)}}\right) \left(R^{\left(\varepsilon\right)}_{h}\left(s\right) + \displaystyle\sum_{s' \in \mathcal{S}} P v^{\left(\varepsilon\right)}_{h+1}\left(s'\right)\right) + v^{\left(\varepsilon\right)}_{h}\left(s\right) \dfrac{\delta}{\lambda^{\left(\varepsilon\right)}} - \displaystyle\sum_{s' \in \mathcal{S}} P_{h}\left(s'|s\right) v^{\left(\varepsilon\right)}_{h+1}\left(s'\right)
\\ &= \left(1 - \dfrac{\delta}{\lambda^{\left(\varepsilon\right)}}\right) R^{\left(\varepsilon\right)} + \left(v^{\left(\varepsilon\right)}_{h}\left(s\right) - \displaystyle\sum_{s' \in \mathcal{S}} P_{h}\left(s'|s\right) v^{\left(\varepsilon\right)}_{h+1}\left(s'\right)\right) \dfrac{\delta}{\lambda^{\left(\varepsilon\right)}}
\\ &= \left(1 - \dfrac{\delta}{\lambda^{\left(\varepsilon\right)}}\right) R^{\left(\varepsilon\right)} + \Delta v^{\left(\varepsilon\right)}_{h}\left(s\right) \dfrac{\delta}{\lambda^{\left(\varepsilon\right)}} ,
\end{aligned} \end{equation}
where we drop the indices $\left(h, s\right)$ as before. Now for any $\lambda < \delta$, we have, for every $i \in \I, j \in \J$,
\begin{equation} \begin{aligned}
& -b \leq R^{\left(\varepsilon\right)}_{ij} \leq b
\\ &\Rightarrow \left(1 - \dfrac{\delta}{\lambda^{\left(\varepsilon\right)}}\right) \left(-b\right) + \dfrac{\delta}{\lambda^{\left(\varepsilon\right)}} \Delta_{ij} v^{\left(\varepsilon\right)} \leq \left(1 - \dfrac{\delta}{\lambda^{\left(\varepsilon\right)}}\right) R^{\left(\varepsilon\right)}_{ij} + \dfrac{\delta}{\lambda^{\left(\varepsilon\right)}} \Delta_{ij} v^{\left(\varepsilon\right)} \leq \left(1 - \dfrac{\delta}{\lambda^{\left(\varepsilon\right)}}\right) b + \dfrac{\delta}{\lambda^{\left(\varepsilon\right)}} \Delta_{ij} v^{\left(\varepsilon\right)}
\\ &\Rightarrow -b + \delta \dfrac{b + \Delta_{ij} v^{\left(\varepsilon\right)}}{\lambda^{\left(\varepsilon\right)}} \leq R'^{\left(\varepsilon\right)}_{ij} \leq b - \delta \dfrac{b - \Delta_{ij} v^{\left(\varepsilon\right)}}{\lambda^{\left(\varepsilon\right)}}
\\ &\Rightarrow -b + \delta \dfrac{b + \Delta_{ij} v^{\left(\varepsilon\right)}}{b - \displaystyle\min_{i' j'} \left| \Delta_{i' j'} v^{\left(\varepsilon\right)} \right|} \leq R'^{\left(\varepsilon\right)}_{ij} \leq b - \delta \dfrac{b - \Delta_{ij} v^{\left(\varepsilon\right)}}{b - \displaystyle\min_{i' j'} \left| \Delta_{i' j'} v^{\left(\varepsilon\right)} \right|}
\\ &\Rightarrow -b + \delta \leq R'^{\left(\varepsilon\right)}_{ij} \leq b - \delta, \text{\;since\;} b + \Delta_{ij} v^{\left(\varepsilon\right)} \geq b - \displaystyle\min_{i' j'} \left| \Delta_{i' j'} v^{\left(\varepsilon\right)} \right| \geq b - \Delta_{ij} v^{\left(\varepsilon\right)},
\\ &\Rightarrow -b + \lambda \leq R'^{\left(\varepsilon\right)}_{ij} \leq b - \lambda,
\end{aligned} \end{equation}
and for any other $\left(i, j\right) \in \mA$,
\begin{equation} \begin{aligned}
& -b + \delta \leq \displaystyle\min\left\{\displaystyle\max\left\{R^{\left(\varepsilon\right)}_{ij}, -b + \delta\right\}, b-\delta\right\} \leq b - \delta
\\ &\Rightarrow -b + \delta \leq R'^{\left(\varepsilon\right)}_{ij} \leq b - \delta
\\ &\Rightarrow -b + \lambda \leq R'^{\left(\varepsilon\right)}_{ij} \leq b - \lambda.
\end{aligned} \end{equation}
In addition, we show that each entry changes by less than $\dfrac{\varepsilon}{2 L H \left| \mathcal{S} \right| \left| \mA \right|}$, for $i \in \I, j \in \J$. In particular, we have
\begin{equation} \begin{aligned}
& \left| R'^{\left(\varepsilon\right)}_{ij} - R^{\left(\varepsilon\right)}_{ij} \right|
\\ &\leq \left| \mQ'^{\left(\varepsilon\right)}_{ij} - \mQ^{\left(\varepsilon\right)}_{ij} \right|
\\ &\leq \left| \left(1 - \dfrac{\delta}{\lambda^{\left(\varepsilon\right)}}\right) \mQ^{\left(\varepsilon\right)}_{ij} + \dfrac{v^{\left(\varepsilon\right)} \delta}{\lambda^{\left(\varepsilon\right)}} - \mQ^{\left(\varepsilon\right)}_{ij} \right|
\\ &= \left| - \dfrac{\delta}{\lambda^{\left(\varepsilon\right)}} \mQ^{\left(\varepsilon\right)}_{ij} + \dfrac{v^{\left(\varepsilon\right)} \delta}{\lambda^{\left(\varepsilon\right)}} \right|
\\ &\leq \left| \dfrac{\delta}{\lambda^{\left(\varepsilon\right)}} \mQ^{\left(\varepsilon\right)}_{ij} \right| + \left| \dfrac{v^{\left(\varepsilon\right)} \delta}{\lambda^{\left(\varepsilon\right)}} \right|
\\ &\leq \left| \dfrac{b H \delta}{\lambda^{\left(\varepsilon\right)}} \right| + \left| \dfrac{b \delta}{\lambda^{\left(\varepsilon\right)}} \right|
\\ &\leq \dfrac{\left(H + 1\right) b}{\lambda^{\left(\varepsilon\right)}} \dfrac{\varepsilon}{L H \left(H + 1\right) \left| \mathcal{S} \right| \left| \mA \right|} \dfrac{1}{2} \dfrac{\lambda^{\left(\varepsilon\right)}}{b} , \text{\;due to the definition of\;} \delta,
\\ &= \dfrac{\varepsilon}{2 L H \left| \mathcal{S} \right| \left| \mA \right|},
\end{aligned} \end{equation}
and for other $\left(i,j\right) \in \mA$,
\begin{equation} \begin{aligned}
& \left| R'^{\left(\varepsilon\right)}_{ij} - R^{\left(\varepsilon\right)}_{ij} \right|
\\ &\leq \left| \displaystyle\min\left\{\displaystyle\max\left\{R^{\left(\varepsilon\right)}_{ij}, -b\right\}, b\right\} - R^{\left(\varepsilon\right)}_{ij} \right|
\\ &\leq \delta
\\ &\leq \dfrac{\varepsilon \lambda^{\left(\varepsilon\right)}}{2 L b H \left(H + 1\right) \left| \mathcal{S} \right| \left| \mA \right|}
\\ &\leq \dfrac{\varepsilon}{2 L H \left| \mathcal{S} \right| \left| \mA \right|}, \text{\;since\;} \lambda^{\left(\varepsilon\right)} \leq b.
\end{aligned} \end{equation}
Therefore we have,
\begin{equation} \begin{aligned}
\ell\left(R^\star\right) - C^\star &\leq \ell\left(R'^{\left(\varepsilon\right)}\right) - C^\star
\\ &\leq \ell\left(R'^{\left(\varepsilon\right)} - R^{\left(\varepsilon\right)} + R^{\left(\varepsilon\right)}\right) - C^\star
\\ &\leq \ell\left(R^{\left(\varepsilon\right)}\right) - C^\star + L \left\|R'^{\left(\varepsilon\right)} - R^{\left(\varepsilon\right)}\right\|_{1}
\\ &\leq \dfrac{\varepsilon}{2} - C^\star + L H \left| \mathcal{S} \right| \left| \mA \right| \dfrac{\varepsilon}{2 L H \left| \mathcal{S} \right| \left| \mA \right|}
\\ &\leq \dfrac{\varepsilon}{2} + L \dfrac{\varepsilon}{2 L}
\\ &= \varepsilon,
\end{aligned} \end{equation}
which concludes the proof.

\textbf{Optimality Gap.} To obtain the result in the linear case, we note that if the cost function is linear, ~\eqref{eq:MGMLP} (restated below) is a linear program (since it does not have the invertibility constraint),
\begin{align}
\displaystyle\min_{R,v,\mQ} & \;\ell\left(R, R^{\circ}\right)
\\ \st & \left[\mQ_{h}\left(s\right)\right]_{\I_{h}\left(s\right)\bullet} \q_{h}\left(s\right) = v_{h}\left(s\right) {\bm 1}_{\left|\I_{h}\left(s\right)\right|}, \forall\; h \in \left[H\right], s \in \mathcal{S}
\notag\\ & \p^\top_{h}\left(s\right) \left[\mQ_{h}\left(s\right)\right]_{\bullet\J_{h}\left(s\right)} = v_{h}\left(s\right) {\bm 1}^\top_{\left|\J_{h}\left(s\right)\right|}, \forall\; h \in \left[H\right], s \in \mathcal{S}
\notag\\ & \left[\mQ_{h}\left(s\right)\right]_{\mAa \setminus \I_{h}\left(s\right) \bullet} \q_{h}\left(s\right) \leq \left(v_{h}\left(s\right)-\iota\right) {\bm 1}_{\left|\mAa \setminus \I_{h}\left(s\right)\right|}, \forall\; h \in \left[H\right], s \in \mathcal{S}
\notag\\ & \p^\top_{h}\left(s\right) \left[\mQ_{h}\left(s\right)\right]_{\bullet \mAb \setminus \J_{h}\left(s\right)} \geq \left(v_{h}\left(s\right)+\iota\right) {\bm 1}^\top_{\left|\mAb \setminus \J_{h}\left(s\right)\right|}, \forall\; h \in \left[H\right], s \in \mathcal{S}
\notag\\ & \mQ_{h}\left(s\right) = R_{h}\left(s\right) + \displaystyle\sum_{s' \in \mathcal{S}} P_{h}\left(s' | s\right) v_{h+1}\left(s'\right), \forall\; h \in \left[H - 1\right], s \in \mathcal{S}
\notag\\ & \mQ_{H}\left(s\right) = R_{H}\left(s\right), \forall\; s \in \mathcal{S}
\notag\\ & \underline{v} \leq \displaystyle\sum_{s \in \mathcal{S}} P_{0}\left(s\right) v_{1}\left(s\right) \leq \overline{v}
\notag\\ & -b+\lambda \leq \left[R_{h}\left(s\right)\right]_{i j} \leq b-\lambda, \forall\; \left(i, j\right) \in \mA, h \in \left[H\right], s \in \mathcal{S} \notag
\end{align}
which we can rewrite it in the standard form for the case when $\iota = \lambda = 0$,
\begin{equation} \begin{aligned}
\displaystyle\min_{x\left(R, v, \mQ\right)} & \ell\left(R, R^{\circ}\right)
\\ & A x = {\bm b},
\\ & x \geq 0,
\end{aligned} \end{equation}
and since $\iota$ and $\lambda$ enters the constraint through ${\bm b}$ linearly, we can write the problem for $\theta = \displaystyle\max\left\{\iota, \lambda\right\}$,
\begin{equation} \begin{aligned}
\displaystyle\min_{x\left(R, v, \mQ\right)} & \ell\left(R, R^{\circ}\right)
\\ & A x = {\bm b}' \coloneqq {\bm b} + \theta {\bm d},
\\ & x \geq 0,
\end{aligned} \end{equation}
for some fixed vector ${\bm d}$.
By ~\cite{bertsimas1997introduction}, in particular, equation $\left(5.2\right)$ in section $5.2$, assuming the optimal solution to ~\eqref{eq:MGMLP} is always finite for every $\iota, \lambda$ satisfying ~\eqref{eq:li}, which is true due to our previous feasibility proof and the fact that the costs are bounded by $b H \left| \mathcal{S} \right| \left| \mA \right|$, we have that the optimal solution can be written as a finite collection of linear functions in the form,
\begin{equation} \begin{aligned}
\ell\left(R, R^{\circ} ; {\bm b}\right) &= \displaystyle\max_{i \in \left[N\right]} y_{i}^\top {\bm b},
\end{aligned} \end{equation}
where $y_{i}$ is the dual optimal solution in a region where $\ell\left(R, R^{\circ} ; {\bm b}\right)$ is linear, and we have,
\begin{equation} \begin{aligned}
\ell\left(R, R^{\circ} ; {\bm b}'\right) &= \ell\left(R, R^{\circ} ; {\bm b} + \theta {\bm d}\right)
\\ &= \displaystyle\max_{i \in \left[N\right]} y_{i}^\top \left({\bm b} + \theta {\bm d}\right)
\\ &= \ell\left(R, R^{\circ} ; {\bm b}\right) + \theta \displaystyle\max_{i \in \left[N\right]} y_{i}^\top {\bm d}
\\ &= \ell\left(R, R^{\circ} ; {\bm b}\right) + O\left(\theta\right).
\end{aligned} \end{equation}
When $\iota = \lambda = 0$, the problem is a relaxation of ~\eqref{eq:GM}, thus we have that the optimal solution to ~\eqref{eq:MGMLP}, denoted by $R'$, satisfies,
\begin{equation} \begin{aligned}
\ell\left(R', R^{\circ} ; {\bm b}\right) &\leq C^\star,
\end{aligned} \end{equation}
and due to our previous feasibility proof, for $\iota, \lambda$ satisfying ~\eqref{eq:li},
\begin{equation} \begin{aligned}
\ell\left(R', R^{\circ} ; {\bm b}'\right) &\geq C^\star,
\end{aligned} \end{equation}
and combined with the previous result,
\begin{equation} \begin{aligned}
\ell\left(R', R^{\circ} ; {\bm b}'\right) &= \ell\left(R', R^{\circ} ; {\bm b}\right) + O\left(\theta\right)
\\ &\leq C^\star + O\left(\theta\right),
\end{aligned} \end{equation}
we have,
\begin{equation} \begin{aligned}
\ell\left(R\left(\iota, \lambda\right), R^{\circ}\right) &= C^\star + O\left(\theta\right).
\\ &= C^\star + O\left(\max\left\{\iota,\lambda\right\}\right).
\end{aligned} \end{equation}

\end{proof}

\subsection{Additional Experiments}\label{app:exp}

\paragraph{Code Details.} We conducted our experiments using standard python3 libraries and the gurobi optimization package. We provide our code in a jupyter notebook with an associated database folder so that our experiments can be easily reproduced. The notebook already reads in the database by default so no file management is needed. Simply ensure the notebook is in the same directory as the database folder (like we have arranged in our uploaded zip). We note that for our benchmark tests, the database was too large to upload directly. Instead we will upload that database on github. However, the scale experiments can be reproduced by using the generation code we included in the notebook.

\paragraph{Classic Two-finger Morra.} Consider the classic Two-finger Morra game. The game's payoff matrix is described in \eqref{game: TFM}. Note that this game is different from the simplified two-finger morra game considered in the main text.
\begin{equation}\label{game: TFM}
    TFM := \begin{pmatrix}
    0 &  2 & -3 &  0 \\
    -2 &  0 &  0 &  3 \\
    3 &  0 &  0 & -4 \\
    0 & -3 &  4 &  0
    \end{pmatrix}
\end{equation}

TFW has infinitely many NEs: each player's strategy can be any convex combination of $(0, 4/7, 3/7, 0)^\top$ and $(0, 3/5, 2/5, 0)^\top$. Since people often naively use uniform mixing, it may be desirable to derive a similar game where uniform mixing is NE. Applying Algorithm~\ref{algo:RAP} with $p = q = (1/4,1/4,1/4,1/4)^\top$ produces the new payoff matrix \eqref{game: TFM-dag}.
\begin{equation}\label{game: TFM-dag}
    TFM^\dagger := \begin{pmatrix}
    0 &  2 & -3 &  0 \\
    -2 &  0 & -2 &  3 \\
    3 &  0 &  0 & -4 \\
    -2 & -3 &  4 &  0 \\
    \end{pmatrix}
\end{equation}

Observe that $TFW^{\dagger}$ is an unfair game with value $-.25$, unlike the original game whose value was $0$. The total cost for the change was $4$.


\paragraph{$5$-action RPSSL.} Consider the generalization of the rock-paper-scissors (RPS) game where each player now has $5$ strategies rock, paper, scissors, spock, and lizard (RPSSL) that we mentioned in the main text. The game's payoff matrix is described in \eqref{game: RPSSL}. Note that this game is different from the 5-action Rock-Paper-Scissor-Fire-Water (RPSFW) game considered in the main text.
\begin{equation}\label{game: RPSSL}
   RPSSL := \begin{pmatrix}
   0 & -1 & 1 & -1 & 1 \\
   1 & 0 & -1 & 1 & -1 \\
   -1 & 1 & 0 & -1 & 1 \\
   1 & -1 & 1 & 0 & -1 \\
   -1 & 1 & -1 & 1 & 0 
   \end{pmatrix}
\end{equation}

Similar to RPS, the unique NE for RPSSL is the uniformly mixed strategy pair $\p = \q = (1/5,1/5,1/5,1/5,1/5)^\top$. Suppose that instead, we wish to skew the distribution to favor the new actions, spock and lizard. Specifically, if $\p = \q = (1/9,1/9,1/9,1/3,1/3)^\top$, running Algorithm~\ref{algo:RAP} produces the new payoff matrix \eqref{game: RPSSL-dag}. 
\begin{equation}\label{game: RPSSL-dag}
    RPSSL^\dagger := \begin{pmatrix}
     0    &     -1    &      1     &    -1    &      1   \\
     1    &      0    &     -1     &     1    &     -1    \\
    -1    &      1    &      0     &    -1    &      1   \\
     1    &     -1    &      1     &     0    &     -1/3 \\
    -1    &      1    &     -1     &     1/3   &   0
    \end{pmatrix}
\end{equation}
We observe the resultant NE is fair with value $0$. The total cost for the change is $1.33$.

\paragraph{Note on Other Cost Functions} For general cost functions, one may use Frank-Wolfe-type algorithms, which call a linear programming (LP) oracle in each iteration. Faster specialized solvers can be used if the cost has additional structures. For example, if the cost function is representable by an SDP or other conic programs, one can use interior methods as implemented in MOSEK or other libraries. If the cost function is quadratic (e.g., squared Euclidean norm), one can use a quadratic program (QP) solver based on the QP simplex methods, such as those implemented in Gurobi. If the cost function is piecewise linear (e.g., L1 norm), one can use an LP solver such as Gurobi or GLPK.

\subsection{Example: Bottled Water}\label{app:water}
\textbf{The original game $R^\circ$}\\
Two bottled water companies, SpringRow and SpringCol, bottle their water from the same natural spring. In doing so, they can choose to let pollution happen (P, e.g. allowing oil to leak from their pumps) or take care not to pollute (NP). If none of them pollute, the water quality is good. If one of them pollutes, the water is bad. If both pollutes, the water is terrible.

SpringRow has a filter system: if the water quality is good or bad, its bottled water product will be satisfactory. But if the water is terrible, it overwhelms the filter and its product is unsatisfactory. SpringCol does not have a filter; its product is only satisfactory if the water is good.

If both company’s products are satisfactory, the consumers are indifferent so each company has 50\% of the market share.
If one company's product is satisfactory but the other's is not, the former takes 100\% of the market share.
If both company's products are unsatisfactory, the consumers complain but have no other choices so again each company has a 50\% market share.
In terms of market share, this is a two-player constant-sum game:
\begin{center} 
\begin{minipage}[c]{.35\linewidth}
\begin{tabular}{|c|c|c|}
\hline
 & P & NP \\ \hline
P & (0.5, 0.5) & (1, 0) \\ \hline
NP & (1, 0) & (0.5, 0.5) \\ \hline
\end{tabular}
\end{minipage}
\end{center}
After shifting and scaling, it shares the same Nash Equilibrium structure as the following two-player zero-sum game:
\begin{center} 
\begin{minipage}[c]{.35\linewidth}
\begin{tabular}{|c|c|c|}
\hline
 & P & NP \\ \hline
P & 0 & 1 \\ \hline
NP & 1 & 0 \\ \hline
\end{tabular}
\end{minipage}
\end{center}
The payoffs can be interpreted as the advantage SpringRow has over SpringCol. We will focus on this zero-sum game as the original game $R^\circ$.

There is something deeply unsettling about the game  $R^\circ$: both players must occasionally and intentionally pollute!
The game has a unique mixed NE: both players play (P, NP) with probability (0.5, 0.5) with a game value of 0.5.
For SpringCol, polluting is self-defense: if it does not pollute, SpringRow will pollute and have a superior product using its filter system.
For SpringRow, polluting is for profit: if it does not pollute, SpringCol will not pollute either but then SpringRow only gets half the market share.

\textbf{The game modification problem}\\
To protect the environment, the government wants both companies to play the pure NE (NP, NP).
The government has the power to change the payoff structure of the game.
For instance, it can detect which company is polluting.
If the government notices a company polluting, it will fine that company $\epsilon$ amount.
Being an altruistic government, it will give that money to the other company.
Then, the modified game $R$ is:
\begin{center} 
\begin{minipage}[c]{.35\linewidth}
\begin{tabular}{|c|c|c|}
\hline
 & P & NP \\ \hline
P & 0 & $1-\epsilon$ \\ \hline
NP & $1+\epsilon$ & 0 \\ \hline
\end{tabular}
\end{minipage}
\end{center}
The payoff for (P, P) remains 0 because both companies are fined but the fines exchange hands.
By inspection as soon as $\epsilon>1$, (NP, NP) will be the unique NE of $R$.
In other words, the government has successfully modified the game to force the two companies into the unique target NE.

But is this the minimal modification the government can do?
Let us formally specify the game modification problem with the tuple $(R^\circ, \p=(0,1), \q=(0,1), b=-\infty, \underline v=-\infty, \overline v=-\infty, \ell)$.
Here, we assumed that there is no bound $b$ on the payoff entries, and the government does not care about the range $[\underline v, \overline v]$ of game value.
It remains for the government to specify its game modification cost function $\ell$ (see section~\ref{sec:lp}).
If the government is concerned with one-time cost, then the above $R$ incurs a cost of $2\epsilon$.
It turns out this is suboptimal: the government could have modified the game as
\begin{center} 
\begin{minipage}[c]{.35\linewidth}
\begin{tabular}{|c|c|c|}
\hline
 & P & NP \\ \hline
P & 0 & $1-\epsilon$ \\ \hline
NP & 1 & 0 \\ \hline
\end{tabular}
\end{minipage}
\end{center}
with $\epsilon>1$ to achieve the same unique pure NE (NP, NP) and only incur $\epsilon$ one-time cost.\footnote{There is no fairness consideration here as to why the government does not fine SpringCol when it alone pollutes.}
But if the government is concerned with forever cost, then either modification is optimal because the target action pair (NP, NP) does not incur a modification cost.

Running our code with all default parameters produces the following game modification solution:
\begin{center} 
\begin{minipage}[c]{.35\linewidth}
\begin{tabular}{|c|c|c|}
\hline
 & P & NP \\ \hline
P & 0 & $-0.01$ \\ \hline
NP & 1 & 0 \\ \hline
\end{tabular}
\end{minipage}
\end{center}
Our code implements one-time modification cost.  The code uses the default value $\iota=0.01$ for the SOW parameter in~\eqref{eq:GMLP}, which ensures a strictly worse switch-out gap.  This is the default when the input argument \textit{value\_gap=None} to our code.
This default value forces $\epsilon$ to be just large enough to produce that gap: in this case $\epsilon=1.01$ and the new entry $R_{P,NP}=-0.01$.
The one-time modification cost is $\ell(R^\circ, R)=1.01$.
The solution produced by our code exactly matches the theoretically predicted behavior.

\end{document}